\documentclass[a4paper,twocolumn,reprint]{article}

\usepackage{marvosym}
\usepackage{amsmath}
\usepackage{graphicx}
\usepackage{color}
\usepackage{array}
\usepackage{arydshln}

\usepackage{wasysym}
\usepackage{multirow}
\usepackage{subfigure}

\usepackage[font={small,it}]{caption}

\author{}

\title{Effect of changing the vocal tract shape on the sound production of the recorder: an experimental and theoretical study.}

\date{}

\begin{document}

\twocolumn[
  \begin{@twocolumnfalse}

\maketitle

\begin{center}
R. Auvray$^{1,2)}$ , A. Ernoult$^{1,2)}$, S. Terrien$^{3)}$, P.Y. Lagr\'{e}e$^{1,2)}$, B. Fabre$^{1,2)}$, C. Vergez$ ^{3)}$
\end{center}
\begin{flushleft}
$^{1)}$ Sorbonne Universit\'{e}s, UPMC Univ. Paris 06, UMR 7190, LAM-Institut Jean Le Rond d'Alembert, 75005 Paris, France. auvray@lam.jussieu.fr\\
$^{2)}$ CNRS, UMR 7190, LAM-Institut Jean Le Rond d'Alembert, 75005 Paris, France.\\
$^{3)}$ LMA, CNRS, UPR 7051, Aix-Marseille Univ, Centrale Marseille, F-13402 Marseille Cedex 20, France.
\end{flushleft}

\begin{abstract}
Changing the vocal tract shape is one of the techniques which can be used by the players of wind instruments to modify the quality of the sound. It has been intensely studied in the case of reed instruments but has received only little attention in the case of air-jet instruments. This paper presents a first study focused on changes in the vocal tract shape in recorder playing techniques.
Measurements carried out with recorder players allow to identify techniques involving changes of the mouth shape as well as consequences on the sound. A second experiment performed in laboratory mimics the coupling with the vocal tract on an artificial mouth. The phase of the transfer function between the instrument and the mouth of the player is identified to be the relevant parameter of the coupling. It is shown to have consequences on the spectral content in terms of energy distribution among the even and odd harmonics, as well as on the stability of the first two oscillating regimes.
The results gathered from the two experiments allow to develop a simplified model of sound production including the effect of changing the vocal tract shape. It is based on the modification of the jet instabilities due to the pulsating emerging jet. Two kinds of instabilities, symmetric and anti-symmetric, with respect to the stream axis, are controlled by the coupling with the vocal tract and the acoustic oscillation within the pipe, respectively. The symmetry properties of the flow are mapped on the temporal formulation of the source term, predicting a change in the even / odd harmonics energy distribution. The predictions are in qualitative agreement with the experimental observations.

\end{abstract}

  \end{@twocolumnfalse}
]

\section{Introduction}

The study of musical instruments based on physical observations has always attempted, to some extent, to include the musician within the sound production mechanisms. The most basic description of the musician control, such as a punctual or a static injection of energy that starts the oscillation, or even a modification of boundary conditions that changes the acoustic properties, already constitute a first attempt to describe the control.

In the case of flute-like instruments (mainly recorders, flutes and organ pipes), the first control parameter that comes to mind  is the blowing pressure. Several studies underlined the importance of other parameters such as the jet velocity (directly related to the blowing pressure), the jet length and height, and the area of the outcoming flow. These parameters, or equivalent dimensionless parameters (Reynolds and Strouhal numbers) are under the controlled of the musician or of the instrument maker depending on the type of instrument considered.

Concerning the blowing pressure, one would intuitively say that it controls the loudness of the sound including spectral enhancement that might occur because of non-linear behaviour. However it has been shown that the variation of blowing pressure also directly affects the frequency of the note and whether the instrument sounds on one register or another.

Understanding the influence of the blowing pressure allows the inclusion of it in sound production models \cite{fletcher1998,Verge1994b,verge1996,DelaCuadra2007,Auvray2012}. These models consider the blowing pressure as a slowly varying control parameter, \textit{i.e.} which varies on time scales much larger than an acoustic period. A recent study justified this assumption by relating ``slow'' variations of the blowing pressure (over 10s) to musical interpretations \cite{Cossette2012}.

On the other hand, the pressure can vary on the same time scales as the acoustic period. During the attack transients, the pressure grows from zero to its final value in ten milliseconds (30ms in very soft attack)\cite{Castellengo1999,Nolle1998,Verge1994b} corresponding to the same order of magnitude as the sound period. The pressure release at the end of the notes and the pressure variations between two slurred notes are of the same time scales. Some models\cite{verge1996,DelaCuadra2007a} attempted to consider these scales of  varying blowing pressure to model transients.

However, as it has already been observed by Verge\cite{Verge1994b}, S\'egoufin\cite{Segoufin2000} and de la Cuadra et al.\cite{DelaCuadra2001,DelaCuadra2008}, the blowing pressure and the resulting jet, are prone to oscillate at time scales much smaller than the ones that can be directly controlled by a human. 
More precisely, the blowing pressure may oscillate at frequencies within the range of the sounding frequency, certainly because of the acoustic coupling between the vocal tract of the musician (or the foot of organ pipe) and the instrument\cite{Auvray2012a}.

A more recent study, led by de la Cuadra et al.\cite{DelaCuadra2008}, revealed major differences between a novice and an experienced flautist considering the blowing pressure, the jet length and height, the area of the outcoming flow. Although it was not used to quantitatively describe the differences between the two players, the authors noticed that the blowing pressure presents more acoustic frequency components in the mouth of the experienced flautist than in that of the novice.No specific observation of the spectral content was carried in this study. Coltman\cite{Coltman1973} led a study of the effect of the mouth resonance on the fundamental frequency in the flute. He showed experimentally that, by varying the volume of an artificial mouth, the fundamental frequency can shift of 10 cents (100 cents = semitone), but the influence on the spectral content is not observed. Chen et al.\cite{Chen2007} led experimental measurements on a recorder player during performance. They found that the recorder player controlled his vocal tract impedance. No evidence of vocal tract tuning with the note played was found. They didn't observe the fluctuating part of the supply pressure. They found that the tongue position influence the broadband component of the sound. They also mentioned an effect on the magnitudes of the first 10 harmonics. 

The effect of the changing of vocal tract shape has been more investigated for reed instruments by\cite{Scavone2008,Chen2009,Chen2011,Chen2012,Freour2013}. A sound synthesis model including the vocal tract has also been proposed by Guillemain\cite{Guillemain2007} for the clarinet. In those studies, the vocal tract is modelled as a series impedance added to the pipe impedance. For flute-like instruments, the coupling through the channel between the instrument and the vocal tract can be written using Bernoulli relation\cite{Verge1994a,Auvray2012}, preventing a linear combination of the impedances of the vocal tract and the instrument.

In the study presented in this paper, the following questions are addressed. Is the musician able to control the acoustic coupling between the recorder and the mouth cavity? What is the effect of this coupling on the sound production ? Lastly, can the existing models be modified to include this effect? This paper is a first step in studying the influence of the vocal tract on sound production of flute-like instruments. While the experiments presented in this paper were carried on recorders, the model presented may be applied to other flute-like instruments.

The paper is structured as follow: section \ref{sec:preliminary_observations} presents a preliminary study with recorder players addressing the first question about the possibility to control the coupling with the vocal tract. A second experiment was developed in laboratory, allowing a better repeatability of the measurement (section \ref{sec:laboratory_experiment}). It used feedback in an artificial playing system to vary the coupling and to observe its effect on sound production.

Data gathered during these studies provided the basis to modify an existing model of sound production,the jet-drive model, by including the effect of the vocal tract. The Jet-Drive model is commonly used to model all flute-like instrument and the modification proposed here may as well be applied to the modelling of all flute-like instrument. The modified model is presented in section \ref{sec:model} while section \ref{sec:discussion} presents the predictions of the model and discusses its limitations.

\section{Preliminary observations}
\label{sec:preliminary_observations}

This section presents preliminary measurements carried out on recorder players in order to investigate the controllability and the effect of the vocal tract shape on the sound production.

\begin{figure}[!t]
\centering
\includegraphics[width=1\linewidth]{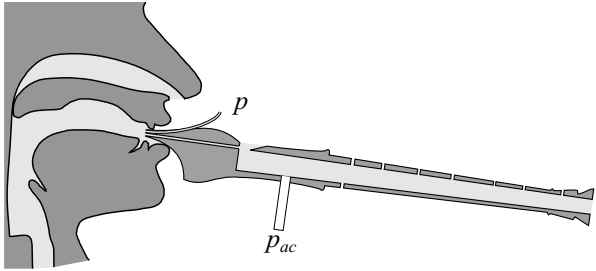}
\caption{\label{fig:flute_micros} Sketch of the modified recorder that allows measurement of the acoustic pressure inside the bore and the pressure inside the mouth.}
\end{figure}

\subsection{Setup and protocol}

Two pressure sensors are mounted on an alto Aesth\'e recorder, made by the recorder maker Jean-Luc Boudreau (figure \ref{fig:flute_micros}). The first sensor is a B\&K microphone model 4938 mounted through the wall, flush with the inner wall and measures the acoustic pressure $p_{ac}$ near the labium. The second sensor is a Honeywell pressure sensor model 176PC14HG1. It allows measurement of the pressure $p$ within the mouth of the player thanks to a capillary tube passing through the mouthpiece and ending in the mouth. The setup followed the ones used by\cite{DelaCuadra2008,Scavone2008,Segoufin2000,Verge1994a,Wilson1996}.

The discussion showed that there is no common playing techniques, shared by professionals players and teachers interviewed, that emphasizes clearly the role of the vocal tract. Without any musical consensus, we could not provide musical instructions that would result in changes of the vocal tract. A first interview with a professional player and with a teacher allowed to establish a protocol for a second interview with five other professional recorder players. 

As it is a preliminary study, the instructions of the second interview focused on isolated notes and musical scales. The recorder players were asked to modify sound features such as the timbre while playing these exercises. Players were given no instructions about using the vocal tract shape but were invited to use whatever techniques they chose to effect the changes in timbre. The freedom given to the players prevents the possibility to carry statistical analysis of data: the data are acquired with different musicians, each of them playing different notes and using different techniques. The aim of these preliminary measurements was to identify the techniques that can be usually used by recorder players.

\subsection{First results}
\label{ssec:prelimineary_observations_first_resuts}
In these interviews, different techniques involving the vocal tract have been mentioned by the musicians. Some of them are, according to the player, directly related to a control of the vocal tract shape (``varying the mouth volume'', ``opening the nasal cavity'', etc.). Others are related to a posture (``gritting the teeth'', etc.). According to the recorder players, these different techniques may be associated to a change in the timbre or other aspects of the sound production such as the ``projection'' or the  ``directivity''.

The fluctuations $p'$ of the supply pressure $p$ are regarded as consequences of the oscillation of the acoustic pressure in the pipe $p_{ac}$. The amplitude of the fluctuating part $p'$ is generally between 10 and 0.1 percent of the steady part. Assuming that the fluctuating part of the supply pressure $p'$ is a small perturbation of the steady part, the coupling can be linearized \cite{Auvray2012a}, and the relation between $p'$ and $p_{ac}$ can be described as a transfer function
\begin{equation}\label{eq:Hp}
\frac{P'(\omega)}{P_{ac}(\omega)} = G_p(\omega) e^{j\varphi(\omega)},
\end{equation}
with $P'$ and $P_{ac}$ the Fourier transform of $p'$ and $p_{ac}$, respectively, and where the $G_p$ and $\varphi$ are the gain and the phase of the transfer function.

\begin{figure}[!t]
\centering
\includegraphics[width=1\linewidth]{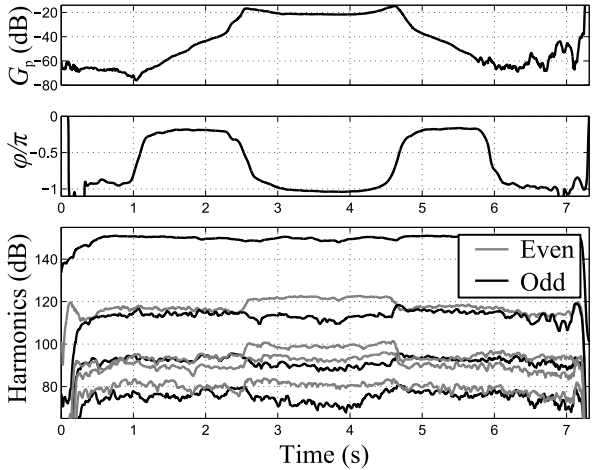}
\caption{\label{fig:mouth_volume} Isolated note performed by a first recorder player (C6, 1046 Hz). He reports changing his mouth volume while playing. From top to bottom: gain of coupling $G_p$, phase of coupling $\varphi$, amplitude of the eight first harmonics of $p_{ac}$ in dB.}
\end{figure}

A change in the vocal tract shape may lead to a change in the gain of coupling. Figure \ref{fig:mouth_volume} shows the gain $G_p$ and the phase of coupling $\varphi$ on an isolated note (C6, 1046 Hz) performed by a first player who reports ``varying his mouth volume''. There is no way to check objectively the movements performed by the recorder player. However, a change in the gain and the phase of coupling has been observed and corresponds to what the recorder player asserts: a narrowing and widening of the mouth. Starting from a ``large'' mouth volume ($1 < t < 2 $ s), the recorder player shrinks it 
($2.5 < t < 4 $ s) then brings it back to the initial volume ($4.5 < t < 5.5 $ s). This can be interpreted as following: as the volume decreases, the mouth impedance at the playing frequency (1047 Hz) rises substantially and the gain of coupling $G_p$ increases from $-$60 dB to $-$20dB. Simultaneously, the phase $\varphi$ decreases from 0 to $\pi$ as the mouth volume decreases.

Varying the vocal tract shape modifies the properties of the coupling. 
The properties of air flow through the channel of the instrument are therefore expected to be modified, which would in turn modify the sound production. The amplitude of the harmonics of the inner acoustic pressure $p_{ac}$ depends on the gain and the phase of coupling (see figure \ref{fig:mouth_volume}). As the gain reaches its maximal value $G_p =-20$ dB ($p'/p_{ac}$=0.1), the even harmonics increase of about 5 dB. The odd harmonics remain however constant, or decrease.

\begin{figure}[!t]
\centering
\includegraphics[width=1\linewidth]{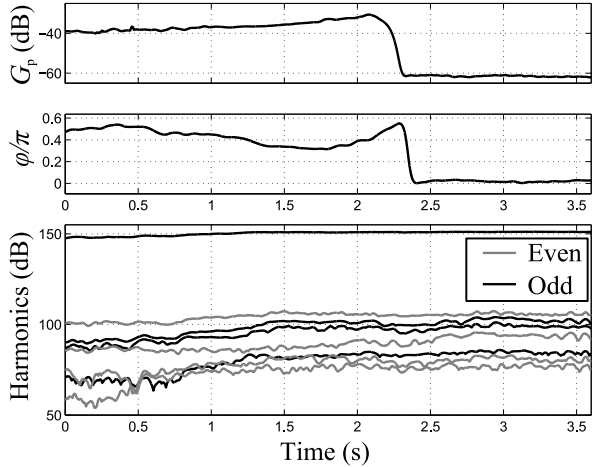}
\caption{\label{fig:nose} Isolated note performed by a second recorder player (G5, 784 Hz). He reports opening his nose cavity gradually. From top to bottom: gain of coupling $G_p$, phase of coupling $\varphi$, amplitude of the eight first harmonics in dB.}
\end{figure}

The change of the spectral content as a function of the coupling with the vocal tract is not systematic. Figure \ref{fig:nose} shows the gain and the phase of coupling and the amplitude of the first eight harmonics of the inner acoustic pressure on an isolated note (G5, 784 Hz) performed by a second recorder player who reports opening his nasal cavity gradually (gradually opening the velum). In this case, there is no monotonous behaviour of the harmonics. This  may be due to a weaker maximal gain than in the previous measurement ($-$30 dB).

This short experimental study confirms that recorder players have the possibility to use on purpose the vocal tract shape as already asserted by Chen et al.\cite{Chen2007}. It may or may not have measurable effect on the harmonics amplitude, depending on the gain of coupling $G_p$, which is a different effect than the one highlighted by the study of Chen\cite{Chen2007} which was focused on the broadband component of the sound.
If the coupling affects the steady part, it is also expected to affect the unsteady part, such as the attack transient. As the stability of the different oscillating regimes of the recorder are sensitive to several parameters, the coupling with the vocal tract may be critical while discussing the stabilities of regimes.

The next section presents an experimental study on the effect of the gain $G_p$ and the phase $\varphi$  of coupling on the sound production.

\section{Laboratory experiments}
\label{sec:laboratory_experiment}

In order to produce reproducible measurements, an artificial mouth for recorder\cite{Ferrand2010} has been set to mimic the acoustic coupling with the vocal tract. One obvious way would be to make different hardware vocal tracts, but it limits the possibility of different coupling which can be investigated. An alternative way is chosen which allows the simulation of much wider coupling conditions with the same setup.  It is based on the injection of an acoustic flux in the artificial mouth, proportional to the acoustic pressure measured within the instrument, using a loudspeaker within the artificial mouth. The experimental setup is detailed in the next section.

\subsection{Experimental setup}

For the playing machine study, the upper (nearly cylindrical) part of a Bressan Zen-On alto recorder is used, and its conical lower joints are replaced by a cylinder of equal diameter, to achieve a near cylindrical instrument, as used in a previous study\cite{Auvray2012}. The total length is 26.5 cm resulting in a resonance frequency of approximately 565 Hz (between a C\#5(554 Hz) and a D5(587 Hz)). Thus the impedance of the resonator is well known, and the comparison with simulated results should be easier. The other parameters of the instrument are the same as an unmodified recorder (window length $W$=4 mm, channel height $h$= 1mm, inner diameter $\diameter$= 19mm) and are thus optimized to produce sound at this frequency which lies in the normal range of the instrument. The artificial mouth consists in a small cavity of diameter 44 mm and height 55 mm yielding a volume $V_0$= 5.6 $10^{-5}$m$^3$. This volume is close to one of the experimenter measured with the method of Coltman\cite{Coltman1973} (measure of the weight of the water contained in the mouth). Air coming from an upstream tank flows into the cavity through a hole of diameter 8 mm. The pressure within the artificial mouth is measured with an Endevco pressure sensor (model 8507C-5). The pressure is controlled by a numerical PID feedback loop implemented with dSpace controller\cite{Ferrand2010}. The effective pressure is compared to the desired one every 40$\mu s$. The calculations needed are performed in this time, which lead to a real time setup. The PID is designed to have a global response time of around 10ms which leads to the filtration of the high frequencies.

The acoustic coupling with the vocal tract is mimicked by the injection of acoustic flux within the cavity. The upstream fluctuations are forced by a feedback loop between the acoustic pressure $p_{ac}$ within the instrument and a loudspeaker Aurasound NSW2 (resonance frequency of 200 Hz, resistance of 6 $\Omega$) set within the artificial mouth. The acoustic pressure within the instrument is measured with an Endevco pressure sensor (model 8507C-2). The feedback loop is managed by the same dSpace controller as the one used for the regulation of the slowly varying pressure. The signal is numerically amplified and phase shifted using a second-order all-pass filter. The parameters of this filter are tuned at the frequency previously measured without forcing. The quality factor of the second-order all-pass filter is equal to 1 while the central frequency is adjusted to produce the desired phase shift at the target oscillating frequency. The modified signal is sent into a power amplifier (Pioneer A107 set with a constant gain) driving the loudspeaker. The artificial mouth and the loudspeaker are held by a larger cavity (volume of 0.13 m$^3$) which purpose is to absorb the backward wave generated by the loudspeaker. The fluctuating part of the supply pressure is checked in order not to affect the regulation of the slowly varying pressure by the PID loop.

The phase and amplitude response of the whole feedback loop depends on the electrical impedance of the loudspeaker, on the lag due to numerical treatment as well as on the response of the instrument itself. The phase and the gain are therefore measured \textit{a posteriori}, after measuring the pressures within the mouth and the instrument. They may differ from what was \textit{a priori} targeted. The phases are estimated with a quadrature phase detection algorithm\cite{Lindsey1972,Auvray2012a} (as used for telecommunication receivers) while the amplitudes are estimated using short term Fourier transform. Results are plotted as functions of the \textit{a posteriori} coupling parameters.


\subsection{Steady sounds and spectral content}

The first of the two blowing conditions used was chosen to study the effect of the coupling on steady sounds, and more precisely on the spectral content of steady sounds. The artificial mouth is supplied with a constant flow. Several runs have been performed for sixteen phase shifts by steps of $\pi/8$ and four coupling gains ($G_p$= $-$34,  $-$26  $-$20,  $-$14 dB) as well as for two mean pressures $\langle p \rangle$= 140 Pa, 400 Pa. These two values are chosen to provide oscillation within the first regime, respectively far from and close to the regime change threshold for increasing pressure regime, respectively far from and close to the regime change threshold for increasing pressure ($\simeq $ 695 Pa).
The former provides a sound close to a pure tone while the latter provides a sound with a strong second harmonic, because of the spectral enhancement of the non-linear saturation due to the increase of pressure.

\begin{figure}[!t]
\centering
\includegraphics[width=1\linewidth]{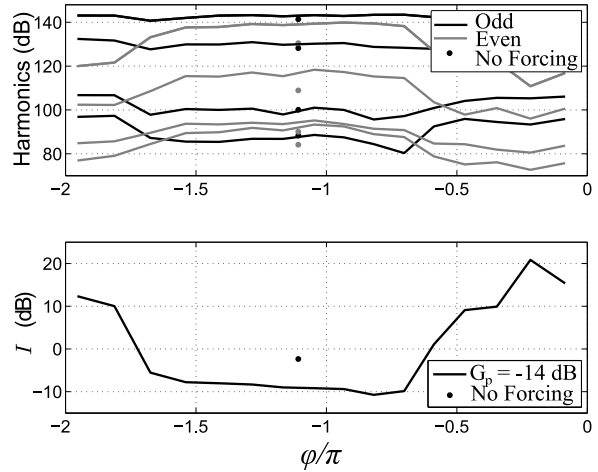}
\caption{\label{fig:expe_harmo_I_vs_phase} Amplitudes of the first eight harmonics and un-parity index $I$ (defined by Eq. (\ref{eq-indice_parite})) as function of the phase of coupling $\varphi$ for the maximal gain ($p'/p_{ac} = 0.2$, $u'/v_{ac} = 0.9$) and for a supply pressure of $\langle p \rangle = 400$ Pa.}
\end{figure}

The spectral content is characterised by the energy distribution discriminating the even and odd harmonics. An \textit{un-parity spectrum} index is defined:
\begin{equation}\label{eq-indice_parite}
I = 10 \log \left( \dfrac{\sum \limits_{p= 1}^{N} \left(a_{2p+1}\right)^2}{\sum \limits_{p= 1}^N \left(a_{2p}\right)^2} \right)  ,
\end{equation}
that weights the sum of the energy of the odd harmonics $\left(a_{2p+1}\right)^2$ over the one of the even harmonics $\left(a_{2p}\right)^2$. By definition, the energy of the fundamental is discarded because it is larger than all the other harmonics, and would tend to smooth the variation of the index $I$. Harmonics from rank 2 to 7 are considered ($N=3$), without the fundamental. Figure \ref{fig:expe_harmo_I_vs_phase} shows the evolution of the first eight harmonics and the associated un-parity index as function of the phase $\varphi$ for the maximal gain $G_p$= $-$14 dB and for the mean supply pressure of 400 Pa. The amplitude of the even harmonics may increase up to 20 dB. It is maximal for a phase $\varphi$ close to $\pi$ and minimal for a phase $\varphi$ close to zero. The amplitudes of the odd harmonics behave in the opposite way. The un-parity index $I$ is a good indicator of the energy distribution between odd and even harmonics. It summarizes in one index the relative evolution of odd and even harmonics. Here, it evolves in a range of 30 dB.

Without coupling, the index is around -2 dB for the two supply pressures. Its value depends, among others, on geometrical parameters, such as the offset between the channel axis and the labium\cite{Fletcher1980}. 

\begin{figure}[!t]
\centering
\includegraphics[width=1\linewidth]{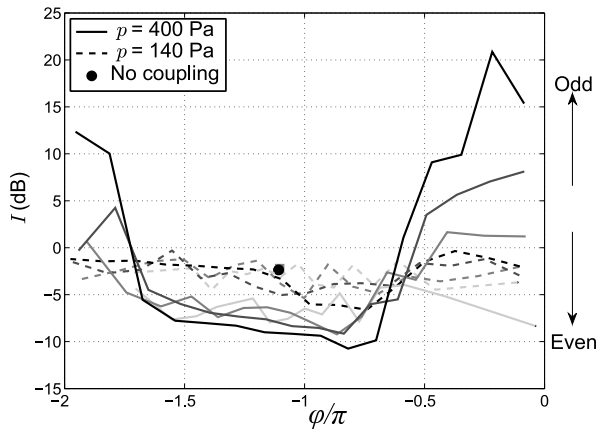}
\caption{\label{fig:expe_I_vs_phase} Un-parity index $I$ (defined by Eq. (\ref{eq-indice_parite})) as function of the phase of coupling $\varphi$ for all gains (increasing gains with darker gray: $G_p$ = $-$14 dB: black, $G_p$ = $-$34 dB: lightest gray) and for two mean supply pressures (140 Pa et 400 Pa).}
\end{figure}

Figure \ref{fig:expe_I_vs_phase} shows the un-parity index for the four gains of coupling and the two supply pressures as function of the phase $\varphi$. The trend described in the previous paragraph for one gain is the same for all gains, at least for the supply pressure of 400 Pa. The un-parity indexes are minimal for phases $\varphi$ close to $\pi$ and maximal for phases $\varphi$ close to zero. The ratio of the odd harmonics over the even increases as the phase gets close to zero and decreases as the phase gets close to $\pi$. However, this effect is less pronounced when the gain of the coupling decreases. For the 140 Pa supply pressure, no clear trend has been observed, even for the maximal gain.

The two supply pressures correspond to two different domains: one with few harmonics, and the other with much more harmonics. The sensitivity of the system with respect to the coupling with the vocal tract seems to depend on the initial energy distribution, \textit{i.e.} without coupling, among the harmonics.

The oscillating frequency results from the balance between the harmonics amplitudes ruled by the non-linear mixing of the exciter. As soon as the relative strengths of the harmonics are changed because of the coupling with the vocal tract, the frequency is expected to deviate from its natural case. For the maximum gain used ($G_p=-14dB$), the frequency varies of around 7 cents between $\varphi$ close to $\pi$ (minimum) and $\varphi$ close to $0$ (maximum). This value is in the same order than the shift found by Coltman\cite{Coltman1973} (around 10 cents).

\subsection{Effects on regime change thresholds}

The second blowing condition aims at studying the effect of the coupling on the stability of the regimes of oscillation. The supply pressure now slowly varies with a triangle shape in time between two extremal values ($\langle p \rangle \in [50, 1200]$ Pa). The range of pressure allows the instrument to sound in its first two oscillating regimes. Several runs have been performed with two phases ($\varphi$= 0 or $\pi$) and three gains of coupling ($G_p$= $-$34, $-$20, $-$14 dB). The phase $\varphi$ is set based on the frequency of the instrument just before the transition to the second register (close to, but above, the resonance frequency of the pipe 565 Hz).

\begin{figure}[!t]
\centering
\includegraphics[width=1\linewidth]{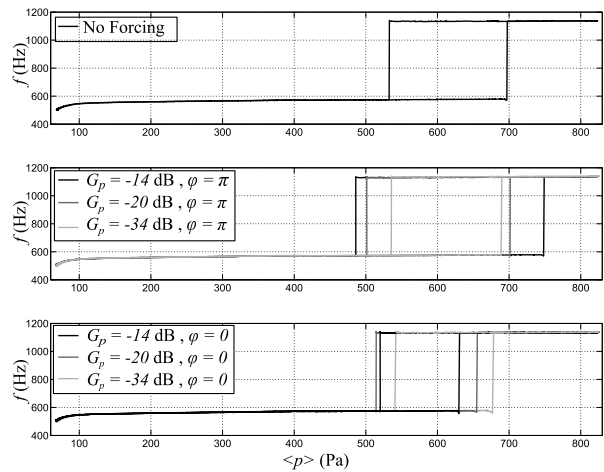}
\caption{\label{fig:expe_seuil} Experimental dimensionless frequency $f/f_1$  as function of the mean supply pressure $\langle p\rangle $ for three coupling conditions. From top to bottom: without coupling, $\varphi = \pi$, $\varphi = 0$.}
\end{figure}

Figure \ref{fig:expe_seuil} shows the oscillating frequency as function of the mean supply pressure $\langle p \rangle$ for all the coupling conditions. The operating range of the first regime is significantly affected by the acoustic coupling with the vocal tract. For a phase $\varphi \simeq \pi$, the increasing threshold from the first to the second regime is shifted from 695 Pa to 745 Pa (7 \%) for the maximal gain ($G_p=-14$ dB). The smaller gains seem to have only a moderate effect on the extent of the first regime. For a phase $\varphi \simeq 0$, the increasing threshold is reduced, on the contrary, from 695 Pa to 630 Pa (6.5 \%), for the maximal gain. For a phase $\varphi \simeq 0$, smaller gains have also an impact on the increasing thresholds.

The decreasing threshold - that from the second to the first regime - is also modified by the acoustic coupling.  For a phase $\varphi \simeq \pi$, the threshold is reduced from 534 Pa to 495 Pa (7 \%), for the maximal gain. For a phase $\varphi \simeq 0$, the threshold is also reduced from 534 Pa to 520 Pa (3 \%), for the maximal gain. The decreasing threshold is reduced for all the coupling conditions.

The phase seems to be the relevant parameter concerning the effect of the changing of vocal tract shape on the stability of the regime, the gain having only an effect of emphasis on threshold shifts. A phase $\varphi$ close to $\pi$ extends the hysteresis range while a phase $\varphi$ close to 0 reduces it.

This result seems counter-intuitive since the phase close to $\pi$ has also been identified as a determining factor in the rise of the even harmonics. One would expect that an increase in the second harmonics within the first regime, close to the frequency of the first harmonic on the second regime,  would induce a reduction of the increasing threshold. This highlights the complexity of the mechanisms that rule the regime changes and calls for deeper investigations.

\section{Model}
\label{sec:model}

The modified model of sound production incorporating changing of vocal tract shape uses three distinct ingredients: the description of the coupling, the growth of instabilities on a jet and the Jet-drive source model that has been shown to predict quite fair results in quasi steady condition\cite{Auvray2012}. The major difficulty lies in combining these descriptions based on different time scales and under different assumptions. In other words the quasi-steady model is enhanced in order to allow fast fluctuations of the blowing pressure.

\subsection{Modelling the coupling}

Most studies related to the effect of the vocal tract tuning focus on the impedance of the vocal tract, described in the frequency domain, and put in series with the instrument impedance \cite{Scavone2008,Chen2009,Chen2011,Chen2012,Freour2013}. This description can not be used in flute-like instruments models. There is no simple relation of continuity between the mouth variables (pressure $p$ and flow) and the acoustic variables into the instrument ($p_{ac}$ and $v_{ac}$). The impedances are not in series. Another description uses, in the temporal domain, a set of two differential equations\cite{Verge1994a} (mass and momentum conservation through the flue channel). 
This offers the advantage to keep the hydrodynamic variables: fluctuations $u'$ of the jet velocity $u = \langle u \rangle + u'$, with $\langle u \rangle$ the steady part of the jet velocity, are induced by the coupling with the vocal tract. As the present study is a first approach in modelling the effect of the coupling, no care will be taken to describe precisely the physical mechanism of this coupling. This mechanism has started to be studied by Auvray et al.\cite{Auvray2012a}.

The description of the coupling in terms of a transfer function between the mouth and the bore pressures (see Eq. (\ref{eq:Hp})) is well suited for the experiments. For the model, an equivalent transfer function can be written between the jet velocity fluctuations $u'$ and the acoustic velocity within the window $v_{ac}$:
\begin{equation}\label{eq:Hu_helmholtz}
\frac{U'(\omega)}{V_{ac}(\omega)} = G_u(\omega) e^{j\phi(\omega)} , 
\end{equation}
with $U'$ and $V_{ac}$ the Fourier transform of $u'$ and $v_{ac}$, respectively. In the present study the gain $G_u$ and the phase $\phi$ are considered as the two control parameters of the coupling. They are varied independently without taking care of their physical origins. It highlights the effects of these two variables on the model.

In the initial model\cite{Auvray2012}, the jet instabilities are triggered by the transverse acoustic velocity near the flue exit. This external perturbation is anti-symmetric and only the anti-symmetric unstable mode is then triggered. This has been confirmed by several experimental observations\cite{Fabre2000}. Taking account of the fluctuating part of the velocity $u'$, another excitation for the jet instabilities is added. The effect of this perturbation on the jet is studied in the next section.

\subsection{Effect of fluctuating velocity on the jet instabilities}
Even if it remains an academic situation, the study of the instabilities of an infinite plane jet for incompressible parallel and non viscous flow still provides insights on the instability mechanisms that occur during flute operation for a finite extent and roughly plane jet.

\begin{figure*}[!t]
\centering
\begin{tabular}{cc}
\includegraphics[width=0.3\linewidth]{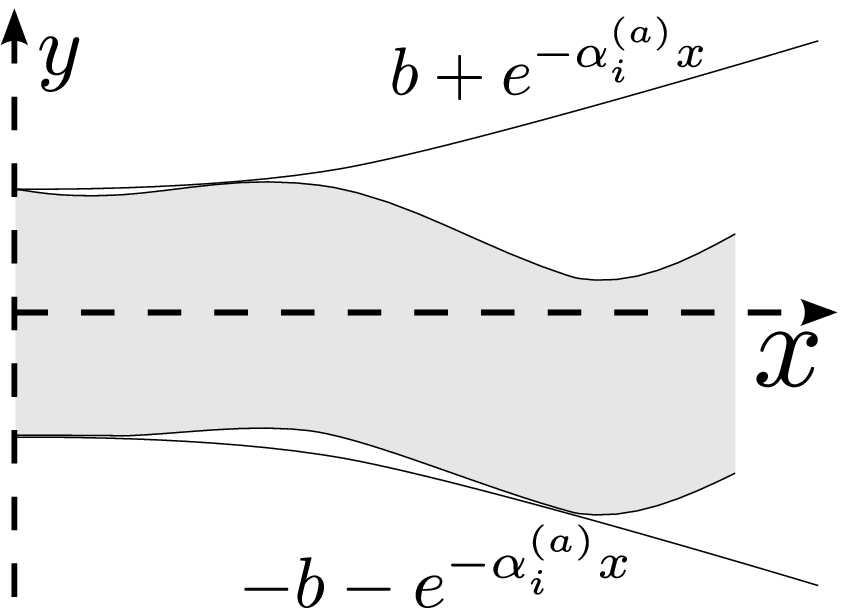} &
\includegraphics[width=0.3\linewidth]{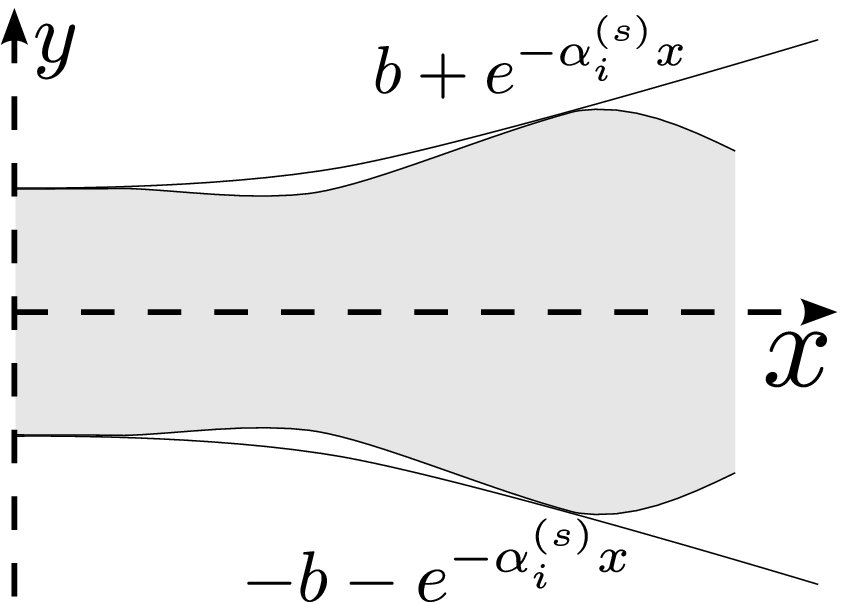}     \\
\includegraphics[width=0.48\linewidth]{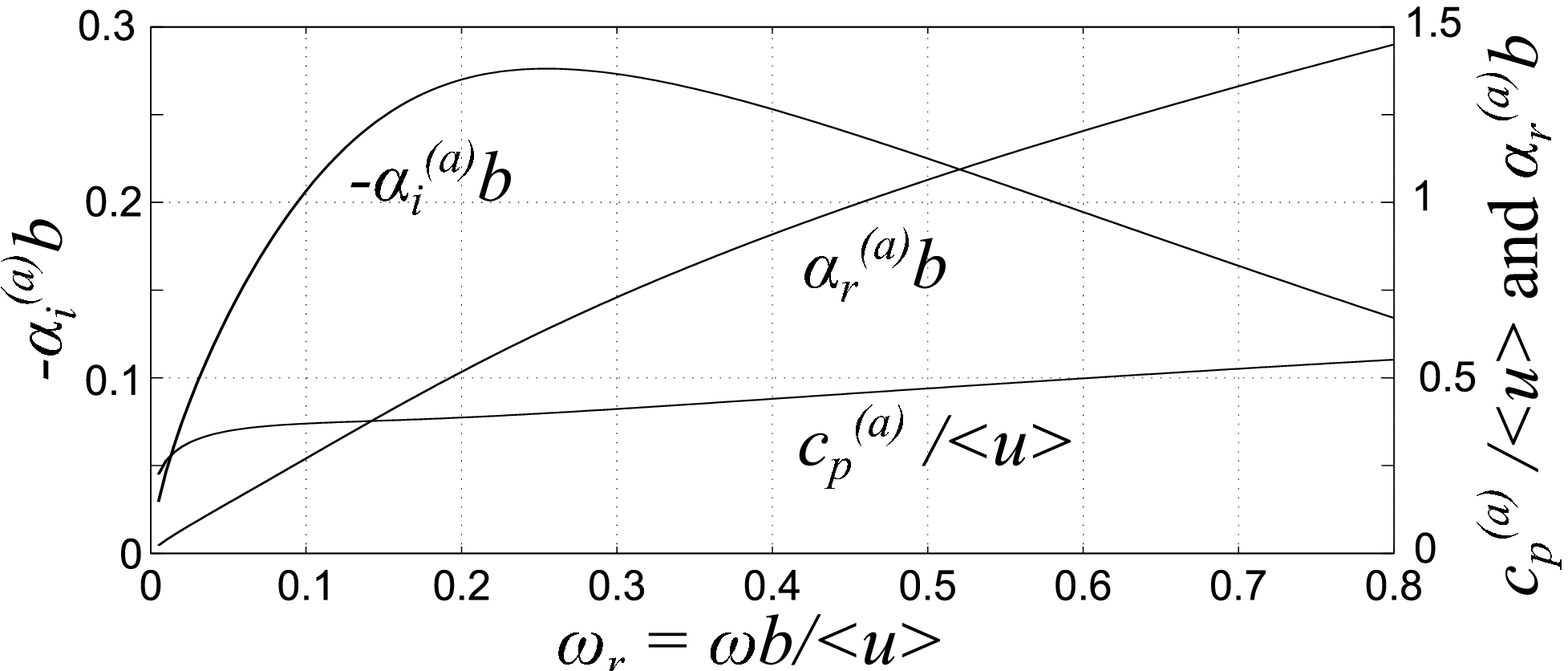}       &
\includegraphics[width=0.48\linewidth]{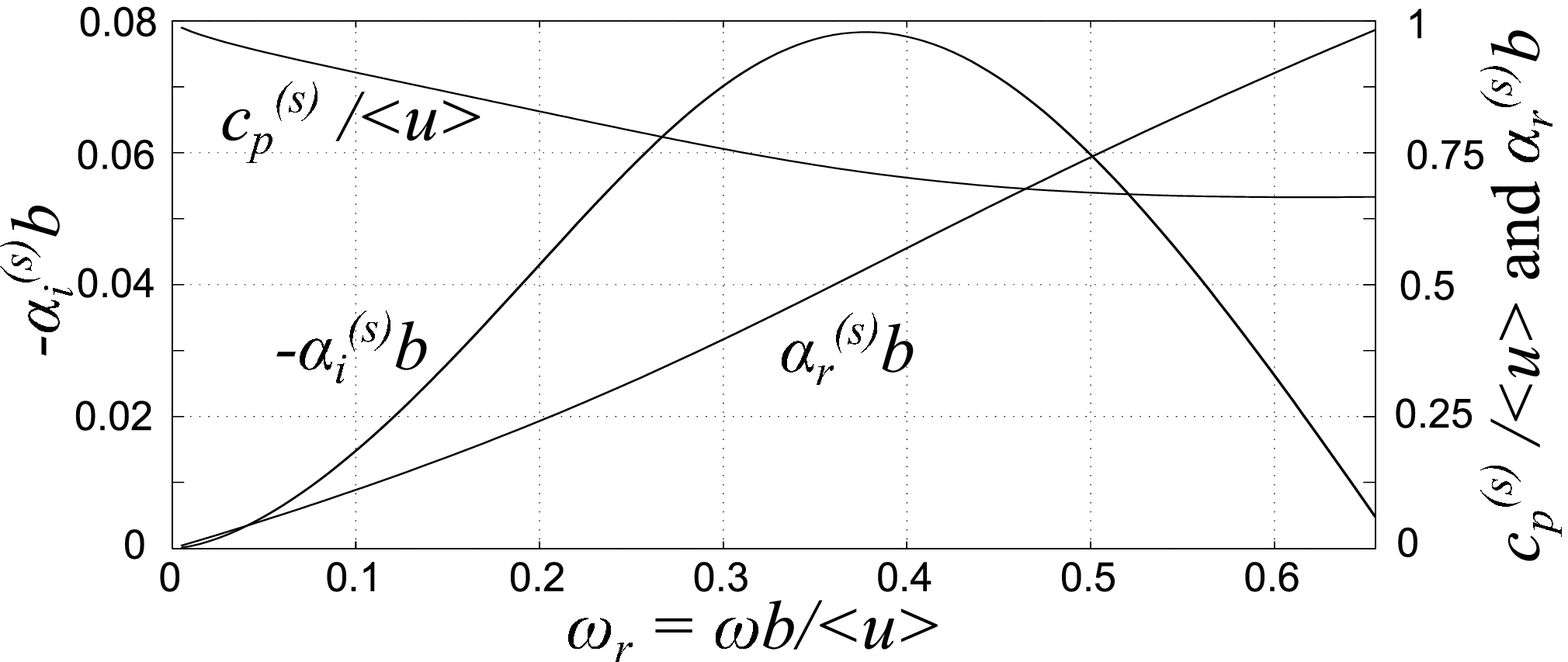}
\end{tabular}
\caption{\label{fig:jet_instabilities} Top: Shape of the perturbation of the jet added to the mean flow for anti symmetric (left) and symmetric (right) cases. Bottom: Corresponding real phase velocity $c_p = \omega/\alpha_r$, real part $\alpha_r$ and imaginary part $-\alpha_i$ of the wave number $\alpha$ as function of the Strouhal number $\omega_r=\omega b/\langle u \rangle$, obtained by numerical solution equation Eq. (\ref{eq:Ray_Mattingly}).}
\end{figure*}

The theoretical jet is assumed to flow along $\mathbf{x}$ direction. The cross-stream profile $U(y)$ along $\mathbf{y}$ direction is assumed to be a Bickley profile:
\begin{equation}
U(y)=\dfrac{\langle u \rangle}{\cosh^2 \dfrac{y}{b} } ~ ,
\end{equation}
where $\langle u \rangle$ is the centerline velocity and $b$ the half-width which is constant in the case of a non viscous flow. A perturbation flow $\mathbf{u} = u \mathbf{x} + v \mathbf{y}$ is added to the mean flow. The linear stability analysis consists in finding a propagative solution of the perturbation field $\mathbf{u} = \hat{\mathbf{u}} e^{i(\omega t-\alpha x)}$. Such a perturbation flow is ruled by the equation of Rayleigh\cite{Mattingly1971}:
\begin{equation}
\label{eq:Ray_Mattingly}
\hspace{-3em} \left[U(y)-\frac{\omega}{\alpha}\right]\left[\frac{d^2\mathbf{v}(y)}{dy^2}-\alpha^2\mathbf{u}(y)\right]-\frac{d^2U(y)}{dy^2}\mathbf{u}(y) = 0,
\end{equation}
where $\omega$ is a real driving pulsation, $\alpha = \alpha_r + i \alpha_i$ the complex wave number and $c_p=\omega/\alpha_r$ the phase velocity. This equation is subject to the boundary conditions:
\begin{equation}
\mathbf{u}'(\pm \infty) = \mathbf{u}(\pm \infty) = 0.
\end{equation}
This set of equations can be solved numerically \cite{Mattingly1971,Nolle1998}. For a symmetric jet profile $U(-y)=U(y)$, Mattingly and Criminale\cite{Mattingly1971} showed that two unstable modes may rise, the antisymmetric $v(-y)=-v(y)$ being more unstable than the symmetric $v(-y)=v(y)$. This is illustrated in figure \ref{fig:jet_instabilities}, which shows the dispersion relations for such symmetric and anti symmetric perturbations. The parameters associated with antisymmetric and symmetric perturbations are respectively indexed by (a) and (s). 

For a given driving pulsation $\omega$ and for a given set of parameters, \textit{i.e.} for a given Strouhal number $\omega_r=\omega b/\langle u \rangle$, the spatial growth factor $-\alpha_i$ describes how the perturbation is amplified while the real phase velocity $c_p=\omega/\alpha_r$ describes at which velocity the perturbation is convected. Without any external excitation, the anti symmetric  perturbation is more unstable than the symmetric, corresponding to the Strouhal number $\omega b/\langle u \rangle \sim 0.25$. As soon as the unstable mode is amplified, the present linear description is no longer valid. The vorticity of the shear layer accumulates at the inflection points. However, the structure of the flow after non-linear development maintains a ``memory'' of the structure presented during the linear development. Thus, the relative position of vortices on the two shear layers of the jet is an indication of which mode has been excited during the initial part. This is confirmed by an experiment of flow visualization presented in appendix \ref{app:visu_sin_vs_var}.

The main hypothesis of the modification of the current simplified model is that the fluctuations of the jet velocity profile due to the acoustic coupling with the vocal tract consist in a symmetric initial perturbation of the jet. The coexistence of two kinds of perturbations, the anti-symmetric acoustic field and the symmetric pulsating jet flow, leads to the linear superimposed growths of the two unstable modes, anti-symmetric and symmetric, with growth factors $\alpha_i$ and phase velocities $c_p$ of order of magnitude predicted by the linear stability analysis of the infinite jet flow. The superimposition is linear during the linear development of the jet.

\subsection{Modified Source model}

As with every self-sustained instrument, the model of sound production of a flute-like instrument is based on a feedback loop system connecting a non-linear exciter and a linear resonator\cite{McIntyre1983}. In the specific case of a flute-like instrument, the model is refined by adding non-linear losses to the former linear acoustic part\cite{Fabre1996}.

The modified model presented in this section is based on the Jet-Drive source model as written by Auvray\cite{Auvray2012}. Only the parts including modifications are extensively presented here (exciter), the others being only mentioned (linear and non-linear acoustics).

The air column of the resonator is driven by a force term written as a pressure difference $\Delta p$ across the window. The acoustic velocity $v_{ac}$ within the window results from the excitation of the resonator by the difference of pressure through the admittance relation:
\begin{equation}\label{eq:Y_source}
Y = \frac{V_{ac}}{\Delta P},
\end{equation}
where $V_{ac}$ and $\Delta P$ are the Fourier transforms of $v_{ac}$ and $\Delta p$, respectively. The resonator part, \textit{i.e.} the admittance, is described as a modal acoustic admittance:
\begin{equation}\label{eq:Y_mod}
Y = \sum_{n = 1}^{\infty} \frac{j \omega Y_n}{\omega_n^2 - \omega^2 + j \varepsilon_n \omega_n \omega},
\end{equation}
where $Y_n$, $\varepsilon_n$ and $\omega_n$ are the amplitude, the damping coefficient and the pulsation of the n-th mode, respectively. The non-linear losses, due to the constriction which yields the formation of a free jet every half acoustic period, are written as a pressure drop across the window:
\begin{equation}\label{eq:DeltaP_losse}
\Delta p_{los} = - \frac12 \rho_0 \frac{v_{ac} \vert v_{ac} \vert}{\alpha_{vc}^2},
\end{equation}
with $\alpha_{vc}=0.6$ a \textit{vena contracta} coefficient. This pressure drop will be added to the source pressure difference.

The exciter part is split into two distinct parts: the birth, growth and convection of the jet instability from the flue exit to the labium and the jet/labium interaction as a mass injection from both sides of the labium.

\smallskip
\noindent \textit{Regarding the antisymmetric perturbations}
\smallskip

Either spreading or not, the jet is described by a unique variable: its center position denoted $\eta(x,t)$. The development of linear anti-symmetric instabilities from the flue exit to the labium is assumed to be well described by the center position of the jet along the same path. As for the perturbation field of the linear stability analysis, the center position is then described as a propagative solution, function of $t-x/c_p^{(a)}$ where $c_p^{(a)}$ is the phase velocity of the anti-symmetric perturbation. The issue of the triggering of the jet perturbation is still an open question, but different authors\cite{Blanc2014,DelaCuadra2005,Fabre2000} agree that the jet is mainly perturbed at the separation point of the flow ($x\simeq 0$) by the acoustic velocity field. Based on flow visualisations, de la Cuadra\cite{DelaCuadra2005,DelaCuadra2007} proposed an expression of the jet center line as a propagative function:
\begin{equation}\label{eq:eta}
\eta(x,t) = e^{-\alpha_i^{(a)} x} \eta_0(t-x/c_p^{(a)}),
\end{equation}
with $-\alpha_i^{(a)} \sim 0.3/h$ the anti-symmetric growth rate, $h$ the height from which the jet emerges and $c_p^{(a)} \sim 0.4 \langle u \rangle$ the anti-symmetric phase velocity and where $\eta_0(t)$ is an ``initial perturbation'' at $x=0$. This expression is only valid a few characteristic distances $h$ downstream so that the initial perturbation $\eta_0$ has no specific physical meaning except as an empirical fit to the data. De la Cuadra found for his experiments:
\begin{equation}\label{eq:eta_0}
\frac{\eta_0(t)}{h} = \frac{v_{ac}(t)}{\langle u \rangle},
\end{equation}
where $v_{ac}$ is the transverse acoustic velocity near the separation point (at $x=0$). The values of the growth factor $\alpha_i^{(a)}$ and the phase velocity $c_p^{(a)}$ vary from one condition to another but remain of this order of magnitude, in agreement with the linear stability analysis presented in the previous section.

\smallskip
\noindent \textit{Adding the symmetric perturbations}
\smallskip

When the velocity profile at the flue exit oscillates because of the acoustic coupling with the vocal tract, symmetric instabilities are expected to arise. As for the anti-symmetric instability that is assumed to modify only the jet center position, the symmetric instability is assumed to modify only the half width of the jet $b$ by adding a perturbation $b'(x,t)$. As for the anti-symmetric case the perturbation is assumed to be described by a propagative solution function of $t-x/c_p^{(s)}$ where $c_p^{(s)}$ is now the phase velocity of the symmetric unstable mode. Similarly, the half width perturbation can be written:
\begin{equation}\label{eq:b_prime}
b'(x,t) = e^{-\alpha_i^{(s)} x} b_0(t-x/c_p^{(s)}),
\end{equation}
with $\alpha_i^{(s)}$ and $c_p^{(s)}$ the growth factor and the phase velocity of the symmetric mode and $b_0(t)$ an ``initial perturbation'' of the jet thickness. The initial perturbation can be written, in first approximation, in the same form as the anti-symmetric perturbation:
\begin{equation}\label{eq:b_0}
\frac{b_0(t)}{h} = \sigma \frac{u'(t)}{\langle u \rangle},
\end{equation}
with $\sigma$ a proportionality coefficient and $u'$ the fluctuating part of the centerline velocity of the jet described in the previous section. The parameter $\sigma$ can be roughly estimated by assuming momentum conservation between an oscillating Poiseuille profile at the flue exit ($x=0$) and a Bickley profile established downstream, as well as conservation of the centerline velocity (see appendix \ref{app:sigma}). The proportionality coefficient $\sigma$ remains a sensitive parameter that further flow visualisations could focus on.

Finally, the total jet velocity profile at the labium is obtained by combining the transverse displacement due to the anti-symmetric perturbation and the modulation of the thickness due to the symmetric perturbation:
\begin{equation}\label{eq:Uw}
U_W(y,t) = \frac{\langle u \rangle}{\cosh^2\dfrac{y-\eta(W,t)}{b+b'(W,t)}}.
\end{equation}

Once it reaches the labium, the jet is assumed to split into two flows $Q_{in}$ and $Q_{out}$ going toward the interior of the instrument and outwards, respectively. Each flux can be split into a steady and a fluctuating components: $Q_{in} = \langle Q_{in} \rangle + Q_{in}'$ and $Q_{out} = \langle Q_{out} \rangle + Q_{out}'$. The flux are assumed to be injected at specific points, behind the labium, and are separated by an acoustic distance $\delta_d=(4\sqrt{2hW})/\pi$, which is also a sensitive parameter of the model\cite{Verge1994a}. This distance being small compared to the acoustic wavelength, the force applied on the air column is due to the acceleration of the small mass of the assumedly incompressible air which is contained between the two injection points. The source term is then written as a pressure difference across the window:
\begin{equation}\label{eq:DeltaP_accel}
\hspace{-3em}  \Delta p_{src} =  -\frac{\rho\delta_d}{WH}\frac{d}{dt} \left( \frac{Q_{in}-Q_{out} }{2} \right),
\end{equation}
with $W$ and $H$ the length and the width of the window, respectively. 
 The difference of fluxes depends on the velocity profile at the labium:
\begin{multline}\label{eq:Qin_integ}
 Q_{in}(t)- Q_{out}(t) = \\ H \left[ \int_{-\infty}^{y_0} U_W(y,t) dy - \int_{y_0}^{\infty} U_W(y,t) dy \right], 
\end{multline}
with $y_0$ the vertical offset between the channel centerline and the labium. The flux $Q_{in}'$ directly depends on the transverse position $\eta$ and the width of the jet $b$ at the labium through Eq. (\ref{eq:Uw}).

Using Eqs. (\ref{eq:Uw}) (\ref{eq:DeltaP_accel}) and (\ref{eq:Qin_integ}) yields the modified source term
\begin{multline}\label{eq:DeltaP_source}
\hspace{-3em} \Delta p_{src}(t) = \\ \frac{\rho \delta_d  \langle u \rangle}{W} \dfrac{d}{dt}\left[ (b+ b'(W,t)) \tanh\frac{\eta(W,t)-y_0}{b+b'(W,t)} \right] .
\end{multline}
The total difference of pressure across the window is given by combining Eqs. (\ref{eq:DeltaP_losse}) and (\ref{eq:DeltaP_source}):
\begin{equation}\label{eq:DeltaP}
\Delta p = \Delta p_{los} + \Delta p_{src} .
\end{equation}

\subsection{Resolution of the model}

While there are several parameters that can be adjusted within the model, we will only focus on those related to the change of vocal tract shape. Some studies propose an investigation of the other parameters\cite{Auvray2012,Coltman2006}.

The control parameters of the simplified model are then the mean or slowly varying blowing pressure $\langle p \rangle$ (or equivalently slowly varying jet velocity $\langle u \rangle$) and the parameters related to the acoustic coupling with the vocal tract. 
 These have been identified to be the gain $G_p$ and the phase $\varphi$ between the fluctuating part $p'$ of the blowing pressure and the acoustic pressure $p_{ac}$ (see Eq. (\ref{eq:Hp})). It is almost equivalent to consider instead the gain $G_u$ and the phase $\phi$ between the fluctuating part $u'$ of the jet velocity and the acoustic velocity $v_{ac}$ (see Eq. (\ref{eq:Hu_helmholtz})). The gain will be varied using the proportionality coefficient $\sigma$ involved in Eq. (\ref{eq:b_0}).

The phases $\phi$ and $\varphi$ are analytically linked. Into the pipe, the acoustic wave is a standing wave. The phase between the acoustic pressure $p_{ac}$ and the acoustic velocity $v_{ac}$ is $\pi/2$. The linearization of the mass conservation between the mouth cavity and the flue exit yields a phase shift of $\pi/2$ between the fluctuating part $p'$ of the blowing pressure and the fluctuating part $u'$ of the jet velocity\cite{Auvray2012a}. These two phase shifts lead to the relation $\phi = \varphi + \pi$.

The numerical resolution is performed by the same algorithm used by Auvray \textit{et al.}\cite{Auvray2014}. It is a step by step time domain resolution. 
The sample rate is known to be sensitive for the numerical procedure\cite{Terrien2012}. The sample rate $Sr$= 0.1 GHz is taken as high as possible while keeping an acceptable computation time.

The auto-oscillation is initiated by injecting a very low amplitude wide band noise. The system then locks on the regime of oscillation corresponding to the blowing condition (mean jet velocity). Two kinds of blowing conditions are provided in order to tackle two issues: one concerning the spectral content, the other the stability of the regimes of oscillation.

For the former, the time of simulation $Ts$ is short (2 s) and the mean jet velocity is constant. The system is solved over several runs for different coupling conditions varying both the gain and the phase of the acoustic coupling.

\begin{table}[!b]
\begin{scriptsize}
\begin{tabular*}{\linewidth}{@{\extracolsep{\fill}} l | l l l }
\hline
\hline
\multirow{3}{1.2cm}{Modal}
            & $\omega_1$= 3547 rad/s       & $\varepsilon_1$= 3.97$~10^{-2}$ & $Y_1$= 1.38$~10^{-3}$ \\
            & $\omega_2$= 2.023 $\omega_1$ & $\varepsilon_2$= 3.18$~10^{-2}$ & $Y_2$= 1.21$~10^{-3}$ \\
            & $\omega_3$= 3.066 $\omega_1$ & $\varepsilon_3$= 2.85$~10^{-2}$ & $Y_3$= 9.81$~10^{-4}$ \\
\hline
\multirow{1}{1.2cm}{Antisym.}
            & $c_p^{(a)}=0.4 \langle u \rangle $       & $\alpha_i^{(a)}=0.3/h$ &   \\
\hline
\multirow{1}{1.2cm}{Symmetric}
            & $c_p^{(s)}=0.8 \langle u \rangle $       & $\alpha_i^{(s)}=0.2/h$ &   \\
\hline
\multirow{5}{1.2cm}{Comput.}
            & $T_s=2$s or $20$s     & $Sr=0.1$GHz &  $c_0$ = 340 m/s \\
            & $\rho_0$=1.2kg/m$^3$  & $\alpha_{vc}=0.6$ &  $y_0=0.1$mm\\
            & $W=4$mm               & $H=12$mm & $h=1$mm  \\
            & $b=2h/5$              &    $\sigma = 0 \leftrightarrow 1$  & \\
            & \multicolumn{3}{l}{$\langle u \rangle = 20$m/s or $30$m/s or $1 \leftrightarrow 56$m/s} \\
\hline
\hline
\end{tabular*}
\end{scriptsize}
\caption{Parameters used for the computation. The modal parameters of the pipe are taken from Auvray \textit{et al.}\cite{Auvray2012}.}
\label{tab:simulation_parameters}
\end{table}

For the latter, the time of simulation is longer (20 s) and the mean jet velocity is slowly varying with a triangle shape in time between its two extreme values (1 m/s and 56 m/s).
Table \ref{tab:simulation_parameters} indexes all the simulation parameters.

%
\section{Simulation results and discussion}
\label{sec:discussion}

\subsection{Simulated steady sounds}

In the simulation, the main control parameter is the phase shift $\phi$ between the fluctuating velocity $u'$ and the acoustic velocity $v_{ac}$. In order to compare the simulation with the experimental data, the results are presented as function of the phase $\varphi = \phi + \pi$ between the fluctuating supply pressure $p'$ and the acoustic pressure $p_{ac}$.

The modal decomposition of the resonator is restricted to three modes in the simulation. As the higher order harmonics are naturally weaker, this has only little impact on the global simulation: only the first harmonics significantly contribute to perturb the jet, at least during the steady state. The un-parity index is approximated, for the simulation, by
\begin{equation}\label{eq:I_simplified}
I \simeq 20 \log \frac{a_3}{a_2},
\end{equation}
with $a_2$ and $a_3$ the amplitudes of the second and third harmonics estimated within the steady part of the simulated sound.

As in the experiments, setting \textit{a priori} the coupling parameters does not allow to predict at which phase and gain the coupling really occurs. In particular, for large gain of coupling, the simulation does not provide stable oscillation on the first regime whereas it was stable without coupling. 

\begin{figure*}[!t]
\centering
\includegraphics[width=0.48\linewidth]{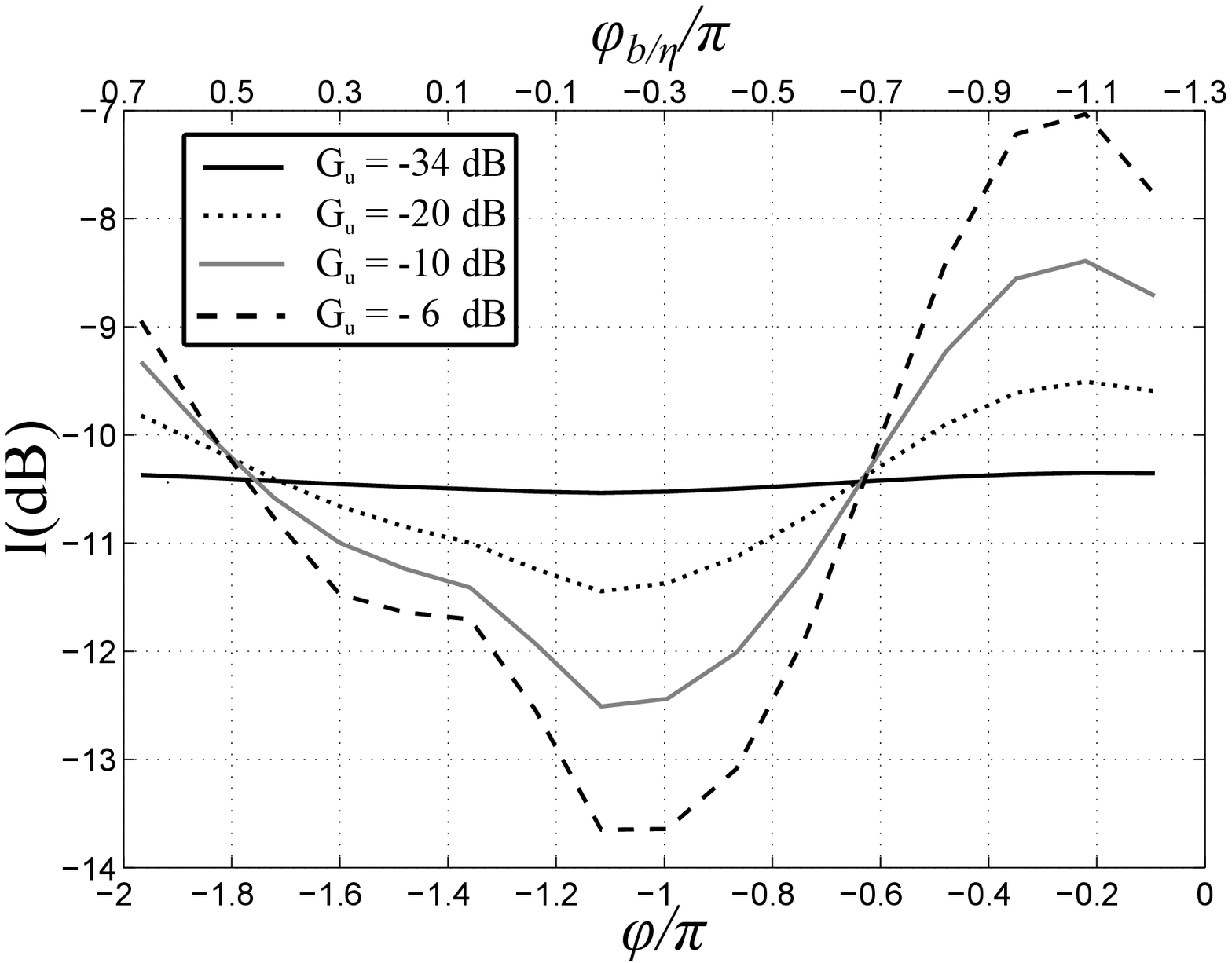} \hfill
\includegraphics[width=0.48\linewidth]{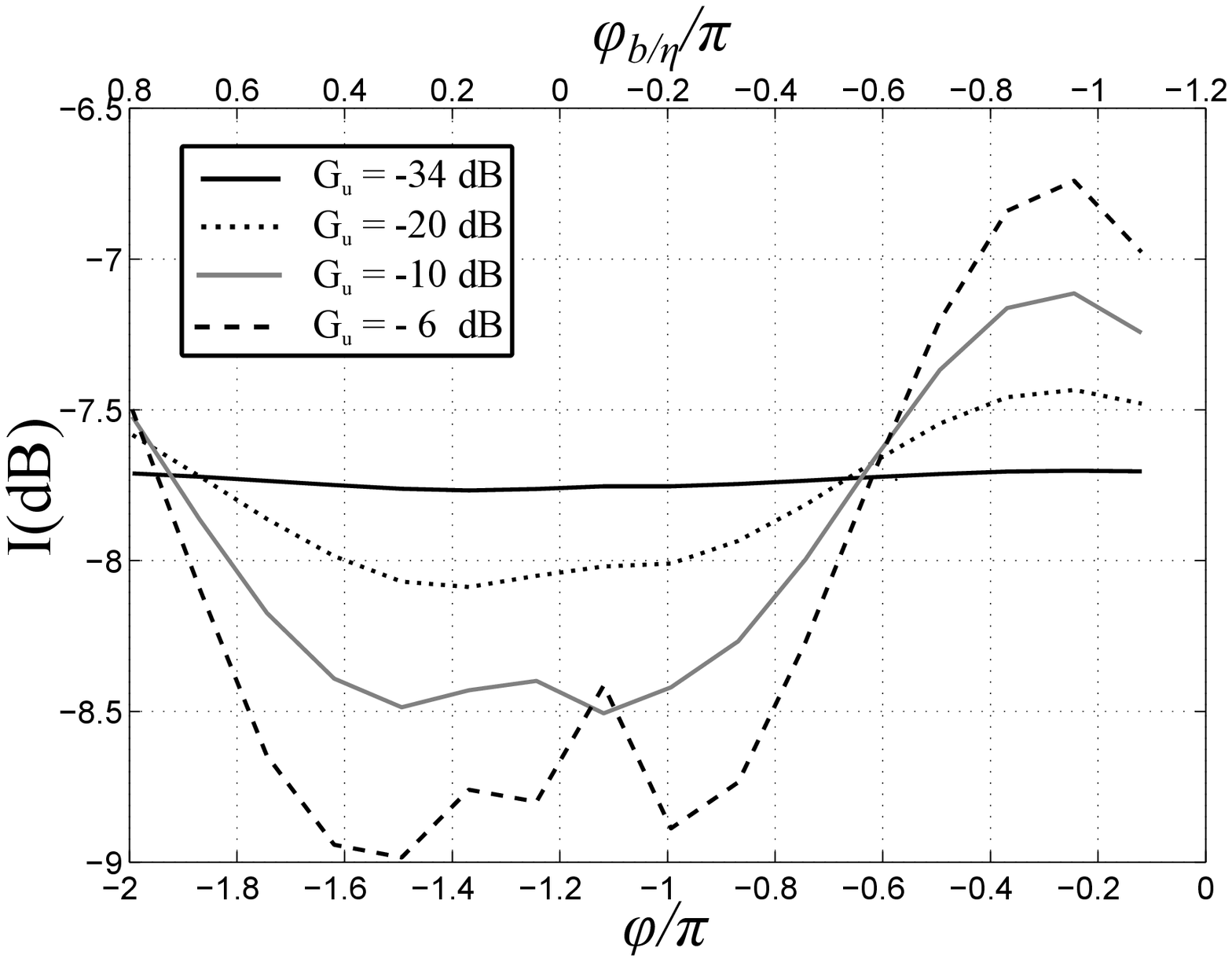}
\caption{\label{fig:simul_I_vs_phase} Simulated un-parity index $I$ as function of the phase of coupling $\varphi$ for several gains of coupling $u'/v_{ac}$ and for two jet mean velocities $\langle u \rangle=20$ m/s (right) and $\langle u \rangle=30$ m/s (left). The phase shift $\varphi_{b'/\eta}$ between the modulation of the jet width $b'$ and the jet center position $\eta$ is also indicated.}
\end{figure*}

The order of magnitude of the un-parity index without coupling is strongly underestimated: $-$10 dB for the simulated index (see figure \ref{fig:simul_I_vs_phase}) when the measured one is $-$2 dB (see figure \ref{fig:expe_I_vs_phase}). 
Nevertheless, the simulation still predicts the trend experimentally observed. Firstly, the variation of the spectral content depends on the phase of coupling $\varphi$. The un-parity index shows minimal value for $\varphi = \pi$ and maximal value for $\varphi = 0$, as in the experiments. The minimum is reached for a phase shift $\varphi_{b'/\eta}$ close to zero.

\begin{figure}[!b]
\centering
\includegraphics[width=1\linewidth]{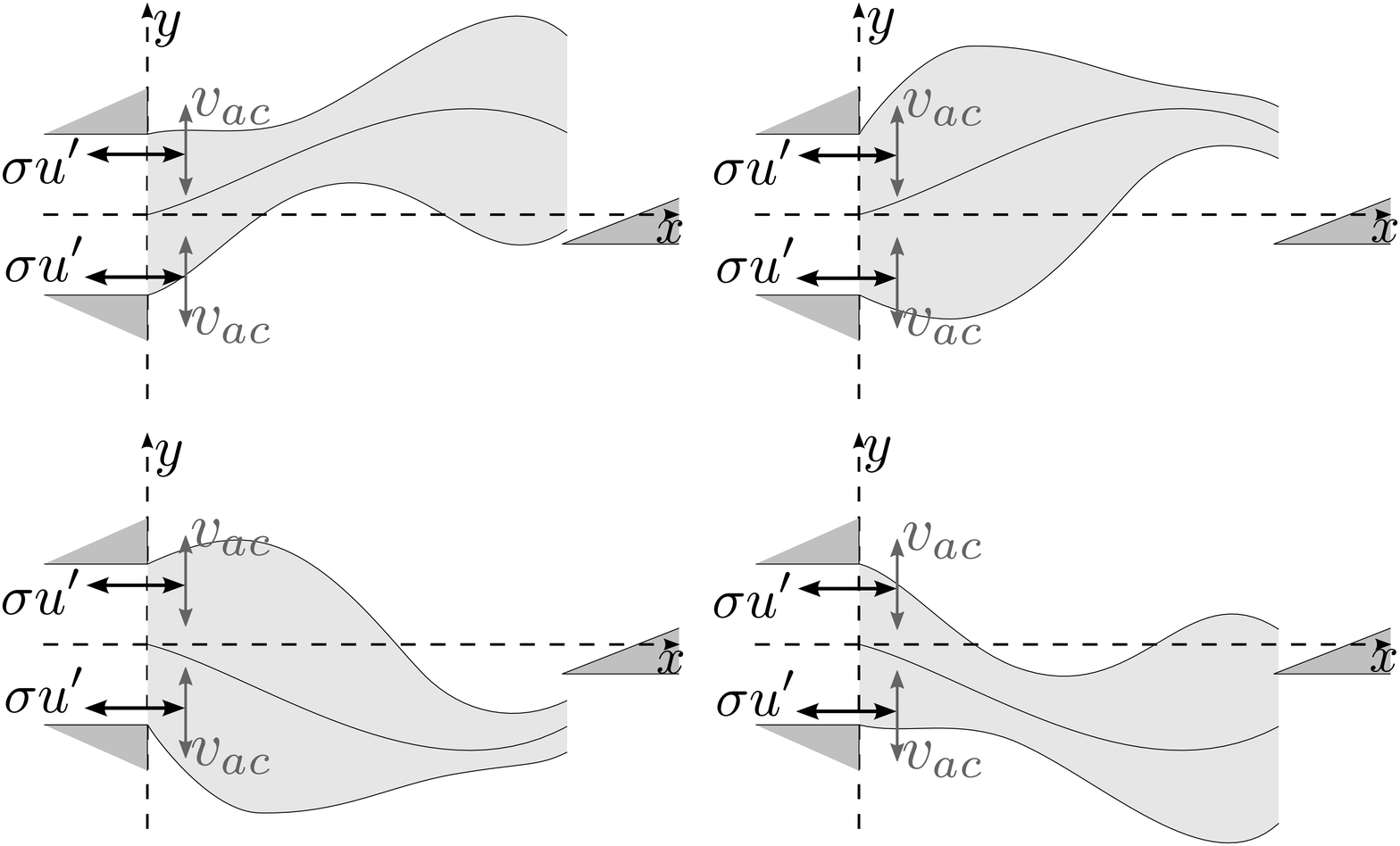}
\caption{\label{fig:schema_jet_drive_modifie} Exaggerated sketch of the superimposition of the anti-symmetric and symmetric modes of jet, excited by the acoustic velocity $v_{ac}$ and the fluctuating jet velocity $u'$, respectively, for two phases of the oscillation (top and bottom) and for two phase shifts $\varphi_{b'/\eta}=0$ (left) and $\varphi_{b'/\eta}=\pi$ (right).}
\end{figure}

This can be intuitively understood by evaluating the flux going toward the interior of the instrument (below the labium) and outwards (above the labium) according to the phase shift $\varphi_{b'/\eta}$. For two oscillations in phase at the labium ($\varphi_{b'/\eta}$=0), as sketched in figure \ref{fig:schema_jet_drive_modifie} (left), the injection of flux above the labium is larger than the one below it. The symmetry of the source is strongly broken. This is the same for two oscillations out of phase (figure \ref{fig:schema_jet_drive_modifie}, right). However, when the oscillations are in quadrature, the effect of the symmetry break is damped: the maximal and minimal thickness of the jet due to the modulation occur when the center position of the jet is zero. In this position, if the offset is zero ($y_0=0$mm),  the problem is symmetrical and the modulation of the thickness have a minimal effect. In the model used, the offset is not zero ($y_0=0.1$mm): the minimal effect of the thickness modulation occurs for $\varphi_{b'/\eta}/\pi \approx$0.7 and 0.6 and not for $\varphi_{b'/\eta}/\pi$=0.5 (see figure \ref{fig:simul_I_vs_phase}).


Secondly, the un-parity index also depends on the gain of coupling, as experimentally observed. The higher the gain of coupling, the more pronounced the effect. The minimum of the un-parity index seems to depend on the jet mean velocity. It is difficult to compare with the experimental data, where it is not the case, at least for the two measured mean pressures. However, the simulation qualitatively agrees with the phase $\varphi$ corresponding to extremal values of the un-parity index and with the effect of the gain of coupling. This calls for a more exhaustive measurement campaign with simultaneous electroacoustic simulated coupling and flow visualisation to investigate the effect of the relative phase $\varphi_{b'/\eta}$ of both symmetric and anti-symmetric instabilities.

\subsection{Simulation of regime change thresholds}

\begin{figure}[t]
\centering
\includegraphics[width=1\linewidth]{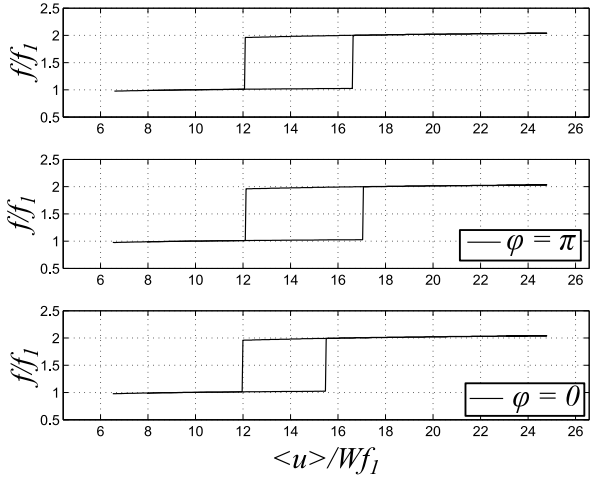}
\caption{\label{fig:simul_seuil} Simulated dimensionless frequency $f/f_1$  as function of the reduced jet velocity $\langle u \rangle /W f_1$ for different coupling conditions. From top to bottom: no coupling, $u'/v_{ac} = 0.1$ and $\varphi=\pi$ and $u'/v_{ac} = 0.1$ and $\varphi = 0$. For clarity purpose, aeolian regimes have been discarded.}
\end{figure}

The Jet-Drive model is known to predict numerous unrealistic aeolian regimes\cite{Auvray2012}. The stability of the oscillating regimes is only discussed regarding the changes between the two main regimes. 
The regime change mechanisms are not fully understood yet\cite{Terrien2014}. The critical parameters of the model have a direct impact on the stability of the different regimes\cite{Auvray2012}.
For small gains of coupling, no variations of the regime change thresholds are predicted by the simulation. However, for large gains, the increasing threshold depends on the phase $\varphi$. Figure \ref{fig:simul_seuil} shows the oscillating frequency as function of the reduced jet velocity $\langle u \rangle/W f_1$ for different phases $\varphi$, when the gain is $G_u$ = $-$20 dB. If the decreasing thresholds are unchanged regarding the coupling conditions, the increasing thresholds do depend on the phase. The prediction qualitatively agrees with the experimental observations: an increase in the hysteresis range occurs for a phase $\varphi$ close to $\pi$ while a reduction occurs for a phase $\varphi$ close to zero.

Every change in the harmonics, in amplitude and phase, seems to have a direct impact on the stability of the regimes. No clear conclusion about the regime change mechanism can be drawn from these results. 
These first simulated results call for another study focusing on the relation between the spectral content and the regime stability, which is out of the scope of the present paper.


\subsection{Limitations of the model}

Mismatches between the predictions and the measurements in terms of spectral content (un-parity index $I$) and regimes stability (regime change thresholds) can be due to a wrong estimation of the numerical values of the parameters of the model as well as the intrinsic assumptions of the model itself.

The set of parameters used here was chosen to match the frequency behaviour (pitch and regime change threshold). Therefore it is natural that discrepancies arise concerning the prediction of the spectral content. The main limitation concerns the number of modes considered. Three modes to describe the admittance may not be enough to describe the filter behaviour of the resonator. The number of modes should be increased and their numerical values should be checked\cite{Terrien2013,Dalmont1995}. Another parameter that has a great impact on the spectral content is the offset $y_0$ between the channel axis and the labium as shown by Fletcher\cite{Fletcher1980} and discussed by Auvray \cite{Auvray2014a}.
Deviation between the real offset and the simulation value $y_0 = b/4$ may be the cause of the discrepancy. The other elements that affect how the jet flow is split at the labium will have an effect on the spectral content of the source. This is the case of the velocity profile, its shape and its thickness. The relation between $h$ and $b$ is not straightforward and requires to know the exact velocity profiles at the flue exit and at a point further downstream. The value of $b$ may differ because the velocity profile at the flue exit may not be a Poiseuille profile and the velocity profile at the labium may not be a Bickley profile. In future work, the new parameters of the model should be varied to make it match the experimental data.

Some limitations of the model may also be due to its strong assumptions.
The description of the hydrodynamics as a one-dimensional pulsating variable may represent a crude assumption compared to a two-dimensional pulsating flow within the channel. However this remains a first order approximation, like the general level of accuracy of the whole model. Furthermore, much care has been taken to describe properly the emerging flow under such excitation.

The description of the jet instability uses a linear theory for infinite plane jet with a symmetric velocity profile. In the case of flute-like instruments, the jet is of finite extent and may not be plane. The velocity profile at the flue exit depends on the history of the flow. In most flute-like instruments, the channel does not show any particular symmetric relation. The profile at the flue exit is not expected to be symmetric. However it may not be totally disordered. The deviation from a symmetric profile can be seen as a slight perturbation of an ideal case. The maximum value of the amplification also depends on the jet velocity profile and thus on the geometry of the channel and also on the way the jet separates from the walls. The amplification is very sensitive to the existence and the shape of chamfers \cite{DelaCuadra2005,Blanc2014}. The present model is however not so refined. The growth factor and the phase velocity remain sensitive parameters the effects of which on the sound production have already been studied\cite{Auvray2012}.

Furthermore, the use of the linear theory has been justified by experimental studies. Flow visualisation experiments estimate the growth $\alpha_i h$ to be in the range [0.1 0.5] for Strouhal numbers $\omega h / 2 \pi \langle u \rangle$ in the range [0 1], with $h$ the height of the channel from which the jet emerges \cite{DelaCuadra2005}. The order of magnitude of the spatial growth factor $-\alpha_i$ and real phase velocity $c_p$ are in qualitative agreement with the linear theory. A slight dependency of the amplitude on the Strouhal number has been found, but on a smaller range of amplitude.

To have more valid values of these parameters, flow visualisation of superimposed symmetric and anti-symmetric instabilities should be conducted. That would provide an experimental estimation of the instabilities parameter. This approximation should be more suitable than a theoretical estimation which relies on strong hypothesis. Furthermore, an electroacoustic coupling experiment could be conducted simultaneously to flow visualisation to address the issue of the relative phase of the jet width and the jet centerline position, which has been shown to be crucial. The energy distribution among the harmonics seems to have an effect on the stability of the oscillating regimes, but the relative phase of the harmonics seems to be more decisive in the stability of the regimes.

\section{Conclusions and perspectives}
\label{sec-conclusion}

This paper represents a study of the effect of changing the vocal tract shape on the sound production of flute-like instruments. The phase of the coupling appears to be the relevant parameter that has an effect on the spectral content in terms of odd/even harmonics energy distribution and on the regime change thresholds. These experimental observations are qualitatively confirmed by the results of simulation using a simplified model of coupling based on the modification of the jet instability.


The identification of the effect of the acoustic coupling allows the development of new protocols in order to study how this control parameter is tuned by recorder player and how this is related to musical intention. This could be performed along with perceptual listening tests to identify the precise points to focus on.

Changing vocal tract shape may also have an important impact during the onset of the oscillation and the attack transient, and therefore on the perception of such musical sounds. The model presented here is based on a steady-state analysis. It is therefore not suited for a study of the transients. All the variations (rise of pressure, fluctuations of pressure and jet velocity due to the acoustic coupling, propagation of the first acoustic wave within the pipe) occurs at the same time scale, the acoustic period. Any attempt to model the coupling during the transient should avoid the description in the frequency domain and should focus on the birth of the jet and the anti-symmetric and symmetric instabilities.

\subsection*{Acknowledgement}
The authors would like to thank the recorder players, teachers and makers Jo\"el Arpin, Philippe Bolton, Cl\'emence Comte, Fran\c{c}oise Defours, Etienne Holmblat, Sarah Lefeuvre and Anne Leleu, who made this study possible.

\appendix

\section{Flow visualisation of symmetric and anti-symmetric ``free'' jet}

\begin{figure*}[!htb]
\centering
\subfigure[$\Delta \varphi \approx 0$. \label{fig:visu_dephas0}]{ \includegraphics[width=0.31\linewidth]{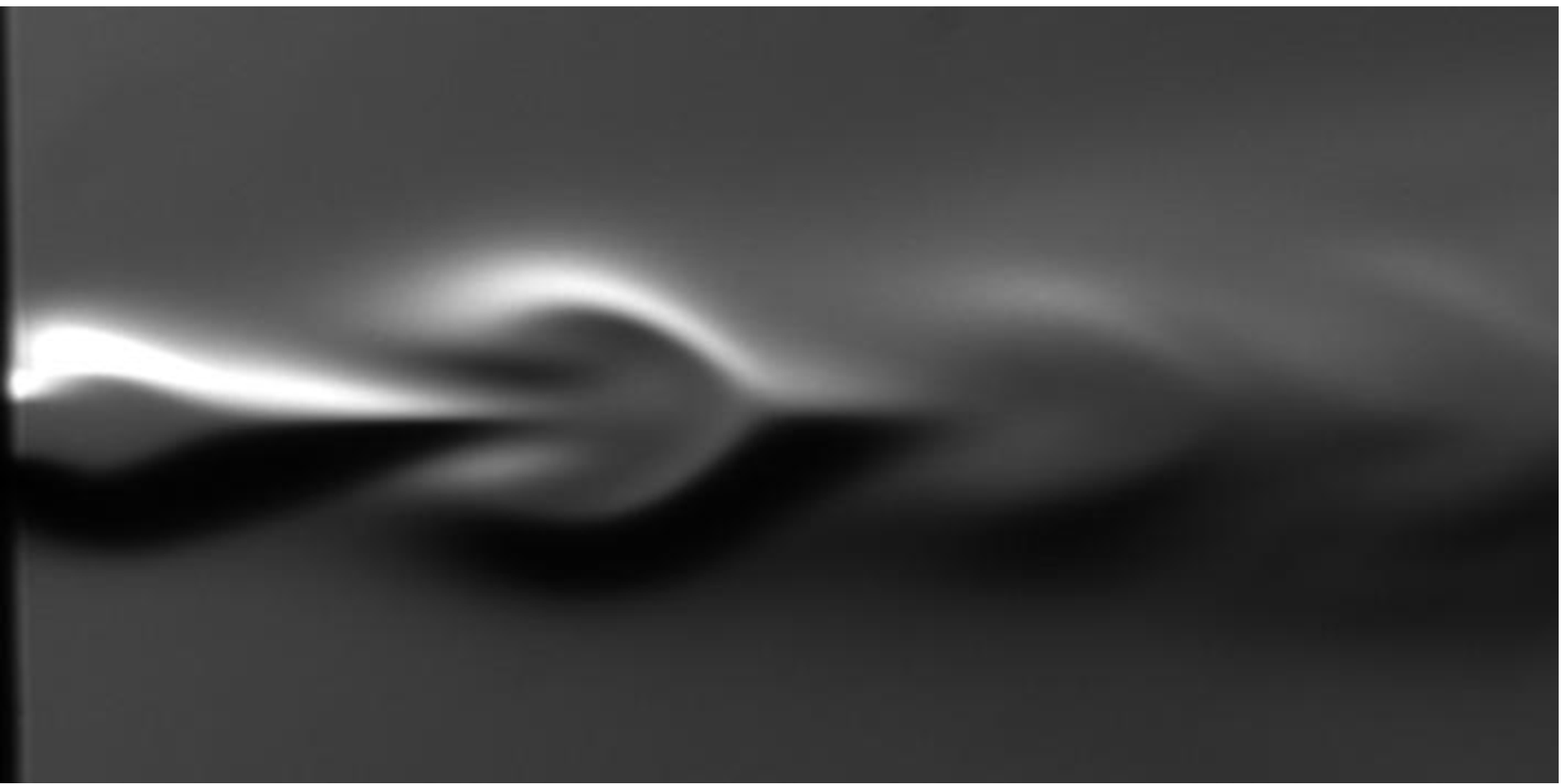} } \hfill
\subfigure[$\Delta \varphi \approx \frac{\pi}{3}$. \label{fig:visu_dephas1}]{ \includegraphics[width=0.31\linewidth]{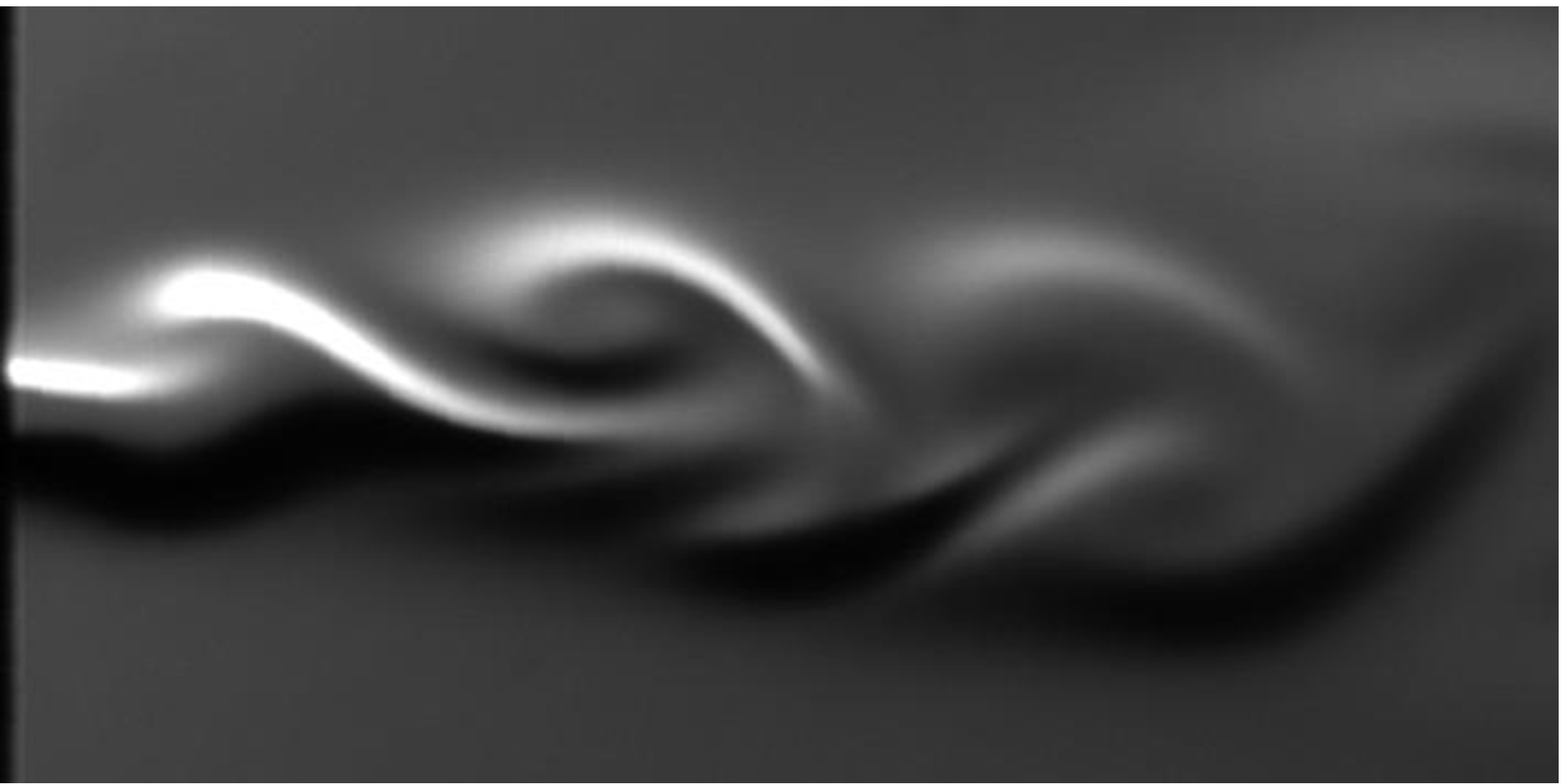} } \hfill
\subfigure[$\Delta \varphi \approx \frac{2\pi}{3}$. \label{fig:visu_dephas2}]{ \includegraphics[width=0.31\linewidth]{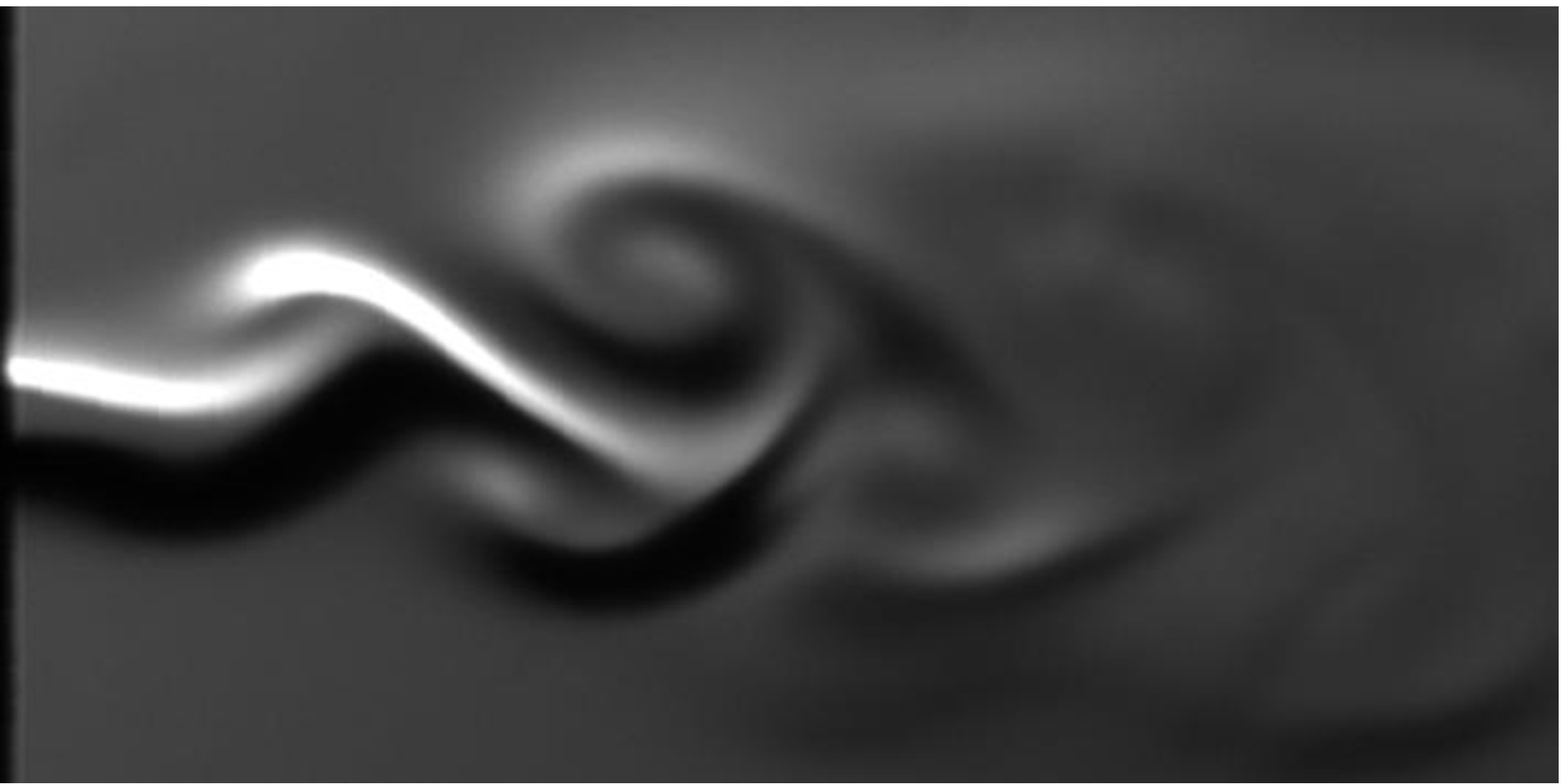}}
\subfigure[$\Delta \varphi \approx \pi$. \label{fig:visu_dephas3}]{ \includegraphics[width=0.31\linewidth]{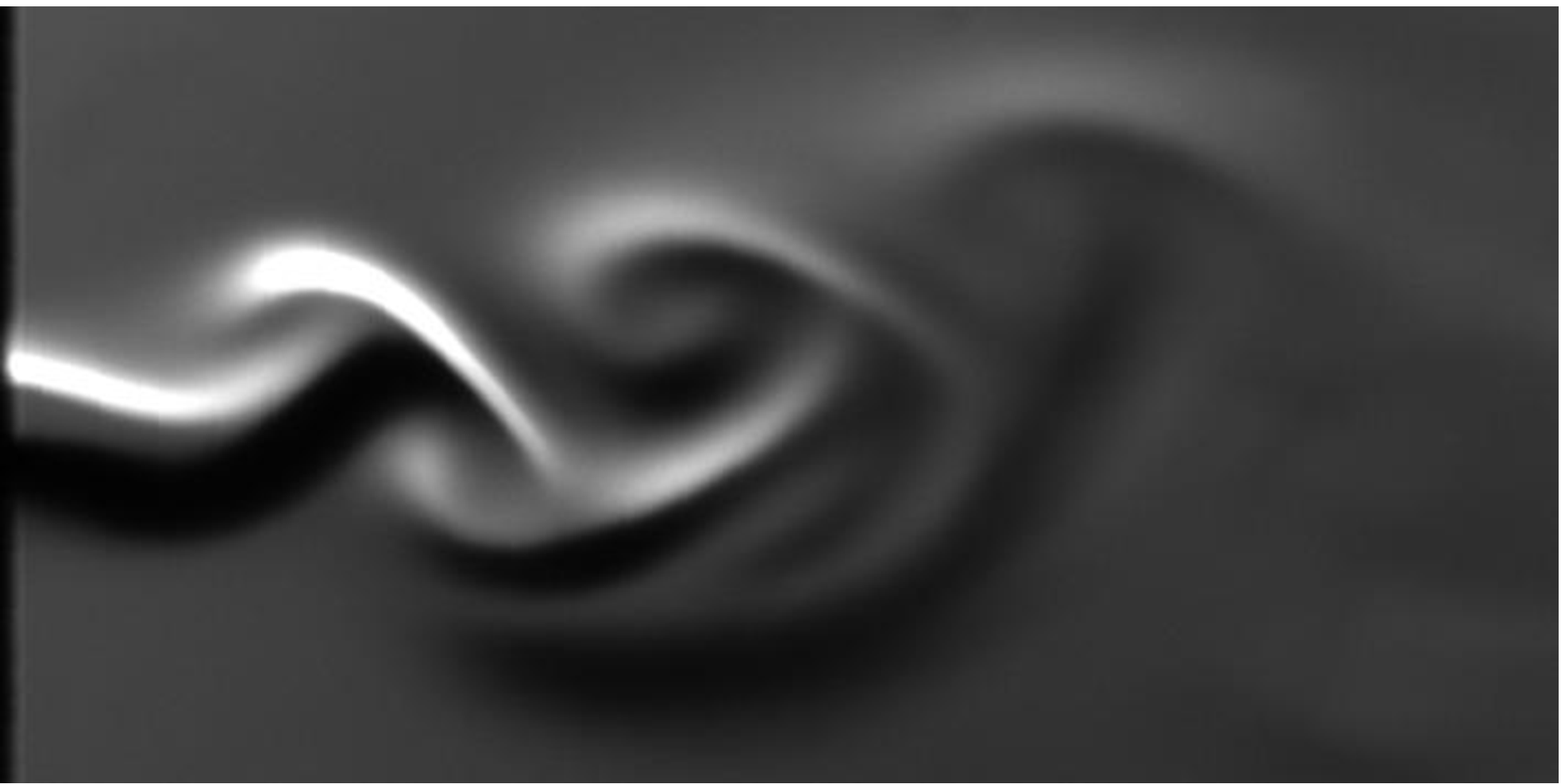} } \hfill
\subfigure[$\Delta \varphi \approx \frac{4\pi}{3}$. \label{fig:visu_dephas4}]{ \includegraphics[width=0.31\linewidth]{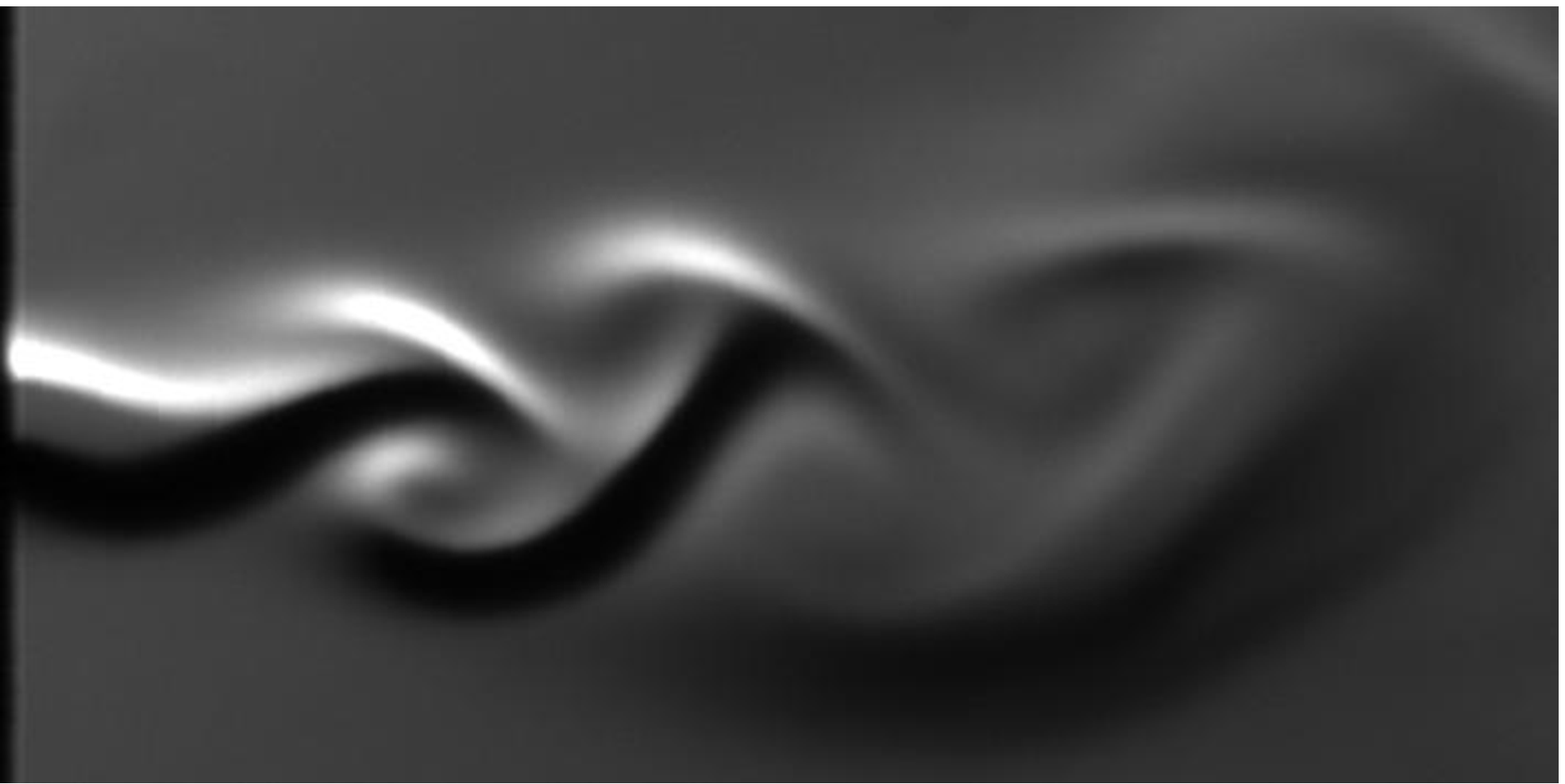} } \hfill
\subfigure[$\Delta \varphi \approx \frac{5\pi}{3}$. \label{fig:visu_dephas5}]{ \includegraphics[width=0.31\linewidth]{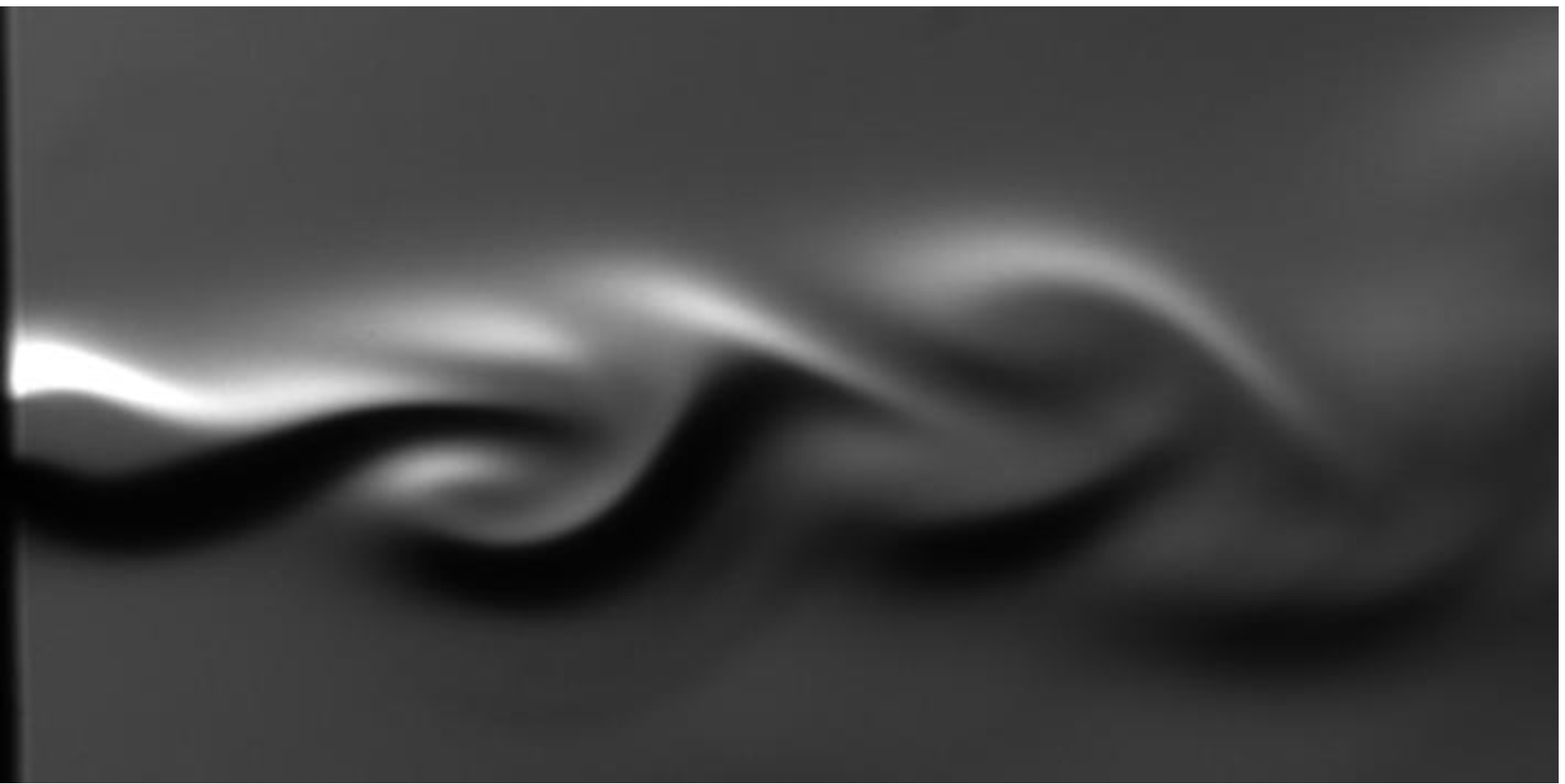}}
\caption{ Flow visualisation of a free jet of jet velocity $\langle u \rangle = 1.5$ m/s excited by two loudspeakers. They are supplied with sinusoidal signals (frequency of 240 Hz) phase shifted by $\Delta \varphi$. The corresponding Strouhal number is $\omega b/ \langle u \rangle \approx 0.4$ with $b=2 h/5$ and $h$ the height of the channel from which the jet emerges. The symmetry property of the excitation is directly related to the phase shift between the two loudspeaker.}
 \label{fig-visue_dephas}
\end{figure*}

This appendix presents a Schlieren flow visualisation experiment of the superimposition of symmetric and antisymmetric jet instabilities.

A pure CO$_2$ jet is released with a centerline velocity of $\langle u \rangle$= 1.5 m/s within an acoustic field driven by two loudspeakers mounted face to face. The experimental set-up is similarly to the one used by De la Cuadra\cite{DelaCuadra2007a}. The symmetry property of the acoustic excitation is gradually varied by gradually phase shifting the two loudspeakers by a value $\Delta \varphi$. The jet is excited at a frequency of 240 Hz.

When the two loudspeakers are in phase ($\Delta \varphi$=0, figure \ref{fig:visu_dephas0}), the acoustic field at the flue channel exit is symmetric, yielding the rise of the symmetric instability only. When the two loudspeakers are out of phase ($\Delta \varphi$=$\pi$, figure \ref{fig:visu_dephas3}), the acoustic field is anti-symmetric, yielding the rise of the anti-symmetric instability only. For transitional phase shifts ($\Delta \varphi$=$\pi/3$, $2\pi/3$, $4 \pi/3$, $5 \pi/3$, figures \ref{fig:visu_dephas1}, \ref{fig:visu_dephas2}, \ref{fig:visu_dephas4}, \ref{fig:visu_dephas5}), the acoustic field symmetry is a combination of both previous cases and both symmetric and anti-symmetric jet instabilities are excited.

The superimposition is linear during the initial linear development of the instabilities, \textit{i.e.} on the first 1/8 of the pictures. When the non-linear behaviour of the instabilities arises, vortices appear on the upper and lower shear layers. Their locations depend on the former state of the jet during the linear development.

\label{app:visu_sin_vs_var}

\section{Estimation of the sensitive parameter $\sigma$}
\label{app:sigma}
It is possible to estimate the parameter $\sigma$, by assuming the momentum conservation between a Poiseuille profile at the flue exit and a Bickley profile established downstream as well as the conservation of the central velocity which gives
\begin{multline}
\int \limits_{-h/2}^{h/2} \rho_0 (\langle u \rangle + u'(t))^2 \left( 1 - \left(\dfrac{y}{h/2}\right)^2 \right)^2 dy = \\ \int \limits_{-\infty}^{\infty} \rho_0 \left( \dfrac{\langle u \rangle}{\cosh^2 (y/b(t))} \right)^2 dy ,
\end{multline}
leading to a relation between the thickness of the Bickley profile and the evolution of the jet velocity
\begin{eqnarray}
b(t) & = & \dfrac{2h}{5} \left( 1 + \dfrac{u'(t)}{\langle u \rangle} \right)^2  \\
& = & \dfrac{2h}{5} + \dfrac{2h}{5}\left(2 \dfrac{u'(t)}{\langle u \rangle} + \left(\dfrac{u'(t)}{\langle u \rangle}\right)^2 \right) . \label{appendix_thickness}
\end{eqnarray}
In Eq. (\ref{appendix_thickness}) the thickness can be seen as a mean value and an "hypothetical initial perturbation"
\begin{equation}
b_0(t) = \dfrac{2h}{5}\left(2 \dfrac{u'(t)}{\langle u \rangle} + \left(\dfrac{u'(t)}{\langle u \rangle}\right)^2 \right) .
\end{equation}
In the case of small perturbation, these expression can be linearised
\begin{equation}
\dfrac{b_0(t)}{h} = \sigma \dfrac{u'(t)}{\langle u \rangle} , 
\end{equation}
where $\sigma = 4/5$. At cost of approximations and assumptions, it is possible to find a linear relation between fluctuation of jet thickness and fluctuation of jet velocity. Even if this reasoning is a good argument for the validity of the linear relation between these two variables, the numerical value of the proportionality coefficient can't be estimated this way and still remains a sensitive point of the model.

\bibliographystyle{plain}

\end{document}